\documentclass[preprint,showpacs,preprintnumbers,amsmath,amssymb]{revtex4}
\usepackage{graphicx}
\begin{document}

%
%
\title{A Master equation approach to line shape in dissipative systems}  
\author{Chikako Uchiyama\(^{1}\), Masaki Aihara\(^{2}\), Mizuhiko Saeki\(^{3,4}\), Seiji Miyashita\(^{3,4}\)}
\affiliation{%
\(^{1}\)Faculty of Engineering, University of Yamanashi,
4-3-11, Takeda, Kofu, Yamanashi 400-8511, JAPAN\\
\(^{2}\)Graduate School of Materials Science,
Nara Institute of Science and Technology\\
8916-5, Takayama-cho, Ikoma, Nara 630-0101 JAPAN\\
\(^{3}\)Department of Physics, The University of Tokyo
7-3-1 Hongo, Bunkyo-ku, Tokyo, 113-0033 Japan\\
\(^{4}\)CREST, JST, 4-1-8 Honcho Kawaguchi, Saitama, 332-0012, Japan 
}%
\date{\today} 
%
\def\ch{{\cal H}}
\def\cha{{\cal H}_{S}}
\def\chb{{\cal H}_{R}}
\def\chab{{\cal H}_{SR}}
\def\co{{\hat {\cal O}}}
\def\cod{{\hat {\cal V}}}
\def\cp{{\cal P}}
\def\cq{{\cal Q}}
\def\cm{{\cal M}}
\def\el{{\cal L}}
\def\chm{{\breve {\cal M}}}
\def\dw{{\dot W}}
\def\ha{{\hat A}}
\def\hb{{\hat B}}
\def\hx{{\hat X}}
\def\hy{{\hat Y}}
\def\hbar{\mathchar'26\mkern -9muh}
\def\rv{{\vec \rho}_{A\;\nu}}
\def\rvm{{\vec \rho}_{A\;-}}
\def\rva{{\vec \rho}_{A}}
\def\pv{{\vec \Psi}}
\def\ih{\frac{i}{\hbar}}
\def\v1{\mathbf{1}}
\def\si{\mathbf{S}_{i}}
\def\sj{\mathbf{S}_{j}}
\def\rij{\mathbf{r}_{ij}}
\def\r12{\mathbf{r}_{12}}
\def\tw{{\tilde \omega}}
\begin{abstract}
We propose a formulation to obtain the line shape of a magnetic response 
with dissipative effects that directly reflects the nature of the environment.
Making use of the fact that the time evolution of a response function is described  by the same equation as the reduced density operator, we formulate a full description of the complex susceptibility.  We describe the dynamics using the equation of motion for the reduced density operator, including the term for the initial correlation between the system and a thermal bath.  In this formalism, we treat the full description of non-Markovian dynamics, including the initial correlation. We present an explicit and compact formula up to the second order of cumulants, which can be applied in a straightforward way to multiple spin systems.  We also take into account the frequency shift by the system-bath interaction.  We study the dependence of the line shape on the type of interaction between the system and the thermal bath. We demonstrate that the present formalism is a powerful tool for investigating various kinds of systems, and we show how it is applied to spin systems, including those with up to three spins.  We distinguish the contributions of the initial correlation and the frequency shift, and make clear the role of each contribution in the Ohmic coupling spectral function. As examples of applications to multispin systems, we obtain the dependence of the line shape on the spatial orientation in relation to the direction of the static field (Nagata-Tazuke effect), including the effects of the thermal environment, in a two-spin system, along with the dependence on the arrangement of a triangle in a three-spin system.
\end{abstract}

\maketitle
\section{Introduction}
\label{sec:1}
Recently, the quantum dynamics of microscopic systems have been observed due to the development of experimental methods. For example, the quantum mechanical magnetization processes of single molecular magnets (SMM) have attracted much interest. Various new aspects of quantum effects are seen in such systems\cite{chiorescu,thomas,bertaina,gatteschi}.

To investigate the energy level structures of SMM molecules,
electron spin resonance (ESR) experiments have been conducted for \({\rm Mn_{12}}\)\cite{hachisuka}.
The quantum tunneling effect was monitored by a proton NMR in \({\rm Fe_{8}}\)\cite{furukawa,ueda} and the dynamics of each magnetic atom was studied using NMR on Mn atoms in \({\rm Mn_{12}}\)\cite{Kuboex}. The temperature dependence of the ESR signal was also studied in \({\rm V_{15}}\)\cite{sakon}.  

Complex susceptibility has been studied for a long time, and the effects of
the exchange and/or dipolar interactions between the numerous
constituent spins have been clarified
\cite{van vleck,Kubo-Tomita,Suzuki-Kubo}.
In order to evaluate the ESR spectra for spatially-structured systems,
theoretical approaches for obtaining line shapes from a microscopic view point 
have recently been proposed by focusing on the effects of the interactions between spins, using direct numerical evaluations of the Kubo formula
\cite{miyaesr,ogasa1,ogasa2,ogasa3}, along with a field theoretical approach\cite{Affleck,Oshikawa}.  In these previous works, the line width comes from the interactions between the spins, which are described by the Hamiltonian system, and the line shape is given by an ensemble of delta-functions. The effects of contact with the thermal bath have not been studied, even though the thermal effect  has attracted interest in studies of microscopic processes. Thus, an approach becomes necessary to introduce the effects of the surroundings, which cause the temperature dependent width of each resonant peak in the complex susceptibility. 

In order to take these effects into account, we have to study an extended system in contact with a thermal bath, and consider the dynamical effects from the thermal bath. For this purpose, the time-evolution of the reduced density operator is usually studied, which is obtained by projecting-out the degrees of freedom of the thermal bath. A standard formalism has been established for the equation of motion for the reduced density operator\cite{Kubo,Louisell,Nakajima,Zwanzig}, which is generally called the quantum master equation. This formalism has been successfully applied to various fields. For example, the natural line width of a two-level (spin) system has been estimated\cite{Louisell,Abragam,Cohen}, and systems with interacting spins\cite{Hamano} and nonlinear spin relaxation\cite{sa} have been studied.  A rapid thermal bath correlation was assumed in these studies, and analyses were therefore made in the Markovian limit.

However, the effect of the finite correlation time of the thermal bath becomes important when we are interested in phenomena where the time scale of the system is comparable to that of the thermal bath. Then, we have to take into
account the time-correlation function of the  thermal bath and the initial condition of the density operator in the above mentioned master equation. 
In the equilibrium state of the total system, which consists of the system, the bath, and the interaction between them, the density operator is not given by a decoupled form.  Therefore, we need to take the contribution from it into account, even though this effect has often been ignored by assuming a factorized form of the density operator.
In the regression theorem\cite{Onsager}, we obtain the time evolution for the average of a quantity for the factorized initial condition and estimate the correlation function from it, which is good in the Markovian limit\cite{Lax,Graham,Haake}.
However, for short time phenomena, we need to estimate the correlation function of the quantity in a non-Markovian evolution, treating the initial correlation correctly\cite{Sassetti,Weiss,Tanimura,Breuer,Saeki}. For example, 
Tanimura\cite{Tanimura} obtained an exact hierarchical formulation with a functional integral for the spectral distribution of an Ohmic form with a Lorentzian cutoff.  
Breuer and Petruccione\cite{Breuer} studied the effects of the initial correlation on the dynamics of a spin-boson system, and pointed out the importance of their contribution. However, they did not obtain an explicit form for the correlation function or the complex susceptibility as a function of the frequency.

In the present paper, we provide a formulation for the complex susceptibility by extending the Nakajima-Zwanzig type of master equation without discarding the non-Markovian effect and the initial correlation. We derive an equation for the motion of the response function.
Then we consider the equation of motion of the quantity \([A, W_{eq}]\), where \(W_{eq}\) is the initial density operator for the total system and \(A\) is a system operator. 
We include the initial correlation between the relevant system and the bath, which is called the ``inhomogeneous term" of the master equation. Since the equation is described by a time-convolution(TC) type of equation for the non-Markovian dynamics, 
the Laplace transformation can be explicitly evaluated. Here we obtain a concrete form of the complex susceptibility. 
It should be noted that the obtained formula is easily evaluated, even in interacting spins, by a concrete numerical calculation.  Moreover, by using the Hilbert-Schmidt (H-S) representation, the formula is compactly expressed.   
In the present formulation, we can include the frequency shift due to 
a system-bath interaction, which comes from the imaginary part of the memory term expressed by the principal value integral of the correlation function of the thermal bath operators.  While we present the formula up to the second order of cumulants, it could easily be extended to the higher orders.

We apply the obtained formula to spin systems linearly interacting with a bosonic bath. For a single spin system, we study the dependence of the line shape on the type of system-bath coupling, e.g., the case of pure dephasing, in which only the diagonal component of the spin interacted with the bath, and the case of longitudinal relaxation, in which the off-diagonal components interacted with the bath (the non-adiabatic interaction). We find that the initial correlation and the frequency shift due to the memory kernel are more dominant in the pure dephasing case than in the non-adiabatic interaction case.  Owing to the usage of the H-S representation, we could extend our formalism to multiple-spin systems in a straightforward way. For a linearly coupled spin chain, the dependence of the peak shift on the angle between the chain and the static field has been studied as the Nagata-Tazuke effect\cite{nagata}. As an example of an application to multispin systems, we study the dependence, including the effects of the thermal environment.  We also study the relationship between the line shape and the geometrical configuration in a three-spin system on a triangle.

This paper is organized as follows: 
We provide a general formulation of susceptibility in Sec. 2.  
The application of the obtained formula to the linear spin-boson model is given in Sec. 3.  Discussions and concluding remarks are given in Sec. 4.

\section{Formulation}
\label{sec:2} 
In this section, we present a formulation of the complex susceptibility of a system in contact with a thermal bath. Generally,
the linear response theory gives the complex susceptibility in the form\cite{Kubo}
\begin{equation}
\chi_{\mu \nu}(\omega)=\lim_{\varepsilon \rightarrow +0} \frac{i}{\hbar} \int_{0}^{\infty} dt \; e^{-i \omega t-\varepsilon t} \; {\rm Tr}[\hb_{\mu}(t),\ha_{\nu}] \rho_{eq} ,
\label{eqn:1}
\end{equation} 
which describes the response of the operator \(\hb_{\mu}\)  to an oscillating external field conjugate to the operator \(\ha_{\nu}\)
with the frequency \(\omega\).  Here, \(\mu\) and \(\nu\) are components of the operators \(\hb\) and \(\ha\), respectively, and \(\rho_{eq}\) denotes an equilibrium state. If we consider the response in a pure quantum state, the time evolution of \(\hb_{\mu}(t)\) is given by \(e^{i{\cal H}_{\rm S}t}B e^{-i{\cal H}_{\rm S}t}\) and \(\rho_{eq}\) is \(e^{-\beta{\cal H}_{\rm S}}/Z_{\rm S}\), where \({\cal H}_{\rm S}\) is the Hamiltonian of the system and \(Z_{\rm S}\) is the partition function of the system at a temperature \(T\) \((\beta=1/k_{\rm B}T)\). On the other hand, to analyze the complex susceptibility under dissipation, we need to describe the time evolution of \(\hb_{\mu}(t)\) by taking into account the interaction between the relevant system and a thermal bath.  

As will be shown in the next section, the dynamics in contact with a thermal bath are not only given by the quantum dynamics of the system, but are also affected by memory effects inherent in the contact with the thermal bath.  The memory effect is often treated in the so-called Markovian approximation\cite{markov}.  This approximation is often used to study the time evolution of the reduced density operator of a system, which leads to the quantum master equation.  As long as the equation has the so-called Lindblad-Kossakowski-Sudarshan form\cite{markov} as in the field of quantum optics, the density operator is positive definite. 
However, it has been pointed out that the Markovian approximation may violate the positivity of the density operator. 
In particular, in a spin-boson model, the breakdown of the positivity has been explicitly reported.  A method to amend this breakdown has been proposed using  a kind of slippage supplement in the initial conditions\cite{pechukas,oppenheim,gaspard}.  Using the Markovian time evolution with this supplement enables us to simulate the time evolution of non-Markovian time evolution, but its validity is limited in a time region larger than the correlation time of the thermal bath\cite{oppenheim,gaspard}.   

Since experimental development has accelerated in recent years, we need to formulate a line shape theory that can correctly describe the non-Markovian effect, including the region of the correlation time of the thermal bath.  
Moreover, the term for the system-bath correlation at an initial time in the equation for the time evolution of the density operator has often been  ignored.  However, the importance of this term has been pointed for obtaining a correct description of the dynamics\cite{pechukas,mori2008}.  Finally, we also need a compact formula that can be evaluated by a concrete numerical method.
For this purpose, we will present a straightforward way to derive a complex susceptibility that includes the initial correlation as well as the non-Markovian effect.

\subsection{Formula of susceptibility}
\label{sec:2a} 
We suppose that a relevant system \(S\) is in contact with a thermal bath \(R\) and that the whole system is in an equilibrium state with temperature \(T\).  Defining the density operator of the whole system as \(W_{eq}\), the linear response theory is extended to give the susceptibility \(\chi_{\mu \nu}(\omega)\) as
\begin{equation}
\chi_{\mu \nu}(\omega)=\lim_{\varepsilon \rightarrow +0} \frac{i}{\hbar} \int_{0}^{\infty} dt \; e^{-i \omega t-\varepsilon t} \; {\rm Tr}_{S+R} [\hb_{\mu}(t),\ha_{\nu}] W_{eq} ,
\label{eqn:2}
\end{equation} 
where \({\rm Tr}_{S+R}\) denotes the trace operation for the whole system.  When we denote \(\cha, \chb, \chab\) as the Hamiltonians of the systems \(S\),\(R\), and the system-bath interaction, the time evolution of an arbitrary operator for the relevant system \(\co\) is determined by the Heisenberg equation,
\begin{equation}
\frac{d}{dt} \co(t)= \frac{i}{\hbar}[\cha+\chb+\chab, \co(t)] \equiv  i \el \co(t).
\label{eqn:3}
\end{equation} 
Defining the total Hamiltonian as \(\ch\) (\(=\cha+\chb+\chab\)), and using the relation as
\begin{eqnarray}
{\rm Tr}_{S+R} [\hb_{\mu}(t),\ha_{\nu}] W_{eq} &=&{\rm Tr}_{S+R} [e^{i \el t}\hb_{\mu},\ha_{\nu}] W_{eq} ={\rm Tr}_{S+R} [e^{\frac{i}{\hbar} \ch t} \hb_{\mu} e^{-\frac{i}{\hbar} \ch t},\ha_{\nu}] W_{eq} \nonumber \\
&& \hspace{-0.5cm}={\rm Tr}_{S+R} \{\hb_{\mu} e^{-\frac{i}{\hbar} \ch t} \ha_{\nu} W_{eq} e^{\frac{i}{\hbar} \ch t} - \hb_{\mu} e^{-\frac{i}{\hbar} \ch t} W_{eq} \ha_{\nu}  e^{\frac{i}{\hbar} \ch t}  \} \nonumber \\
&&\hspace{-0.5cm}={\rm Tr}_{S+R} \hb_{\mu} \{e^{-i \el t}[\ha_{\nu},W_{eq}]\},
\end{eqnarray} 
we can rewrite Eq.~(\ref{eqn:2}) in the form,
\begin{eqnarray}
\chi_{\mu \nu}(\omega)&=&\lim_{\varepsilon \rightarrow +0} \frac{i}{\hbar} \int_{0}^{\infty} dt \; e^{-i \omega t-\varepsilon t} \;{\rm Tr_{S+R}} \hb_{\mu} \{e^{-i \el t}[\ha_{\nu},W_{eq}]\}   \nonumber \\
&=&\lim_{\varepsilon \rightarrow +0} \frac{i}{\hbar} \int_{0}^{\infty} dt \; e^{-i \omega t-\varepsilon t} \; {\rm Tr_{S}} \hb_{\mu} {\rho_{A}}_{\nu}(t) ,
\label{eqn:4}
\end{eqnarray} 
with
\begin{equation}
{\rho_{A}}_{\nu}(t) \equiv {\rm Tr_{R}} e^{-i \el t} [\ha_{\nu}, W_{eq}].
\label{eqn:5}
\end{equation}
where  \({\rm Tr_{R}}\) denotes the trace operation over the thermal bath.
With the Fourier-Laplace transform \(f[\omega]= \int_{0}^{\infty} dt e^{-i \omega t} f(t)\) where \(f(t)\) is an appropriate function,
we find that the susceptibility \(\chi_{\mu \nu}(\omega)\) is given by
\begin{equation}
\chi_{\mu \nu}(\omega)= \frac{i}{\hbar} {\rm Tr_{S}} \hb_{\mu} {\rho_{A}}_{\nu}[\omega] = \chi_{\mu \nu}'(\omega)-i \chi_{\mu \nu}''(\omega). \label{eqn:6}
\end{equation} 

The above formulation shows that the procedure to obtain the complex susceptibility reduces to obtaining \({\rho_{A}}_{\nu}[\omega]\).  As shown in Appendix \ref{sec:a1}, the time evolution of \({\rho_{A}}_{\nu}(t)\)  is given in a form of a `` master" equation by using the projection operator technique.  Here, we define the projection operator to be \(\cp=\rho_{R} {\rm Tr_{R}}\).  Up to the second order of the system-bath interaction \(\chab\), we have 
\begin{equation}
\frac{d}{dt} {\rho_{A}}_{\nu}(t) = - \frac{i}{\hbar} [\cha, {\rho_{A}}_{\nu}(t)] +
      \int_{t_{0}}^{t} d \tau \Xi_{2}(t-\tau) {\rho_{A}}_{\nu}(\tau)+ \Psi_{2}(t),
\label{eqn:7}
\end{equation}
where the kernel \(\Xi_{2}(t)\) and the inhomogeneous term \(\Psi_{2}(t)\) are given by 
\begin{equation}
\Xi_{2} (t-\tau) =  {\rm Tr_{R}} (-i \el_{1}) e^{-i \el_{0} (t-\tau)} \cq (-i \el_{1}) \rho_{R} , 
\label{eqn:8} 
\end{equation}
and
\begin{equation}
\Psi_{2}(t)= {\rm Tr_{R}} (-i \el_{1}) e^{-i \el_{0} t} \cq [\ha_{\nu}, - \rho_{0} \int_{0}^{\beta} d\lambda \ch_{1}(-i \hbar \lambda)],
\label{eqn:9} 
\end{equation} 
respectively.  In Eqs.~(\ref{eqn:8}) and (\ref{eqn:9}), we used the following definitions \(\el_{k} \co = \frac{1}{\hbar}[\ch_{k}, \co]\), and \(\ch_{0} = \cha+\chb\) and \(\ch_{1} \equiv \chab \) .  Because the whole system is assumed to be in an equilibrium state, we have to take into account the contribution of the initial correlation between the system and the thermal bath, which is represented by the inhomogeneous term \(\Psi_{2}(t)\).

From Eq.~(\ref{eqn:6}) and Eq.~(\ref{eqn:7}), the susceptibility \(\chi_{\mu \nu}(\omega)\) is given by
\begin{eqnarray}
\chi_{\mu \nu}(\omega)
&=& \frac{i}{\hbar} {\rm Tr_{S}} \hb_{\mu} \frac{1}{i \omega + i \el_{S} - \Xi_{2}[ \omega]} ({\rho_{A}}_{\nu}(0) + \Psi_{2}[\omega]),
\label{eqn:10}
\end{eqnarray} 
where we define \(i \el_{S} \co=\frac{i}{\hbar} [\ch_{S}, \co]\) with an arbitrary operator \(\co \).  Our remaining task is to obtain \(\Xi_{2}[ \omega]\) and \(\Psi_{2}[\omega]\).  For this purpose, we give concrete expressions for  \(\Xi_{2}(t)\) and \(\Psi_{2}(t)\) in the next subsection. 

\subsection{Concrete expressions for \(\Xi_{2}(t)\) and \(\Psi_{2}(t)\)}
\label{sec:2b} 
For simplicity, we consider the case in which the interaction between the system and the thermal bath is given in the form
\begin{equation}
 \ch_{1} = \chab \equiv \hbar \hx \, \hy 
\label{eqn:11n} 
\end{equation} 
with the system operator \(\hx\) and the thermal-bath operator \(\hy\).  In this form, the second and third terms in Eq.~(\ref{eqn:7}) are given by
\begin{eqnarray}
\int_{0}^{t} d \tau \Xi_{2}(t-\tau) {\rho_{A}}_{\nu}(\tau) \nonumber \\
&& \hspace{-4cm}=(-\frac{i}{\hbar})^2 \int_{0}^{t} d \tau {\rm Tr}_{R} [\ch_{1},[\ch_{1}(-(t-\tau)), e^{-i \el_{0}(t-\tau)} {\rho_{A}}_{\nu}(\tau)] \nonumber \\
&& \hspace{-4cm}=- \int_{0}^{t} d \tau  
[ \Phi(t-\tau) \hx \hx(-(t-\tau)) \{e^{-i \el_{S} (t-\tau)} {\rho_{A}}_{\nu}(\tau)\} \nonumber \\
&&\hspace{-1.5cm}-\Phi(t-\tau) \hx \{e^{-i \el_{S} (t-\tau)} {\rho_{A}}_{\nu}(\tau)\} \hx(-(t-\tau))\nonumber \\
&&\hspace{-1.5cm}-\Phi(-(t-\tau)) \hx(-(t-\tau)) \{e^{-i \el_{S} (t-\tau)} {\rho_{A}}_{\nu}(\tau)\} \hx \nonumber \\
&&\hspace{-1.5cm}+\Phi(-(t-\tau)) \{e^{-i \el_{S} (t-\tau)} {\rho_{A}}_{\nu}(\tau)\} \hx(-(t-\tau)) \hx], \nonumber \\
\label{eqn:12n} 
\end{eqnarray} 
and
\begin{eqnarray}
\Psi_{2}(t)&=&(-\frac{i}{\hbar}) {\rm Tr}_{B} [\ch_{1},[\ha_{\nu}(-t), -\rho_{A} \int_{0}^{\beta} d\lambda \hx(-i \hbar \lambda-t)] \nonumber \\
&=&i \int_{0}^{\beta} d\lambda \Phi(-i \hbar \lambda-t) \{
 (\hx \ha_{\nu}(-t) \rho_{A}  \hx(-i \hbar \lambda-t)  - \hx \rho_{A} \hx(-i \hbar \lambda-t) \ha_{\nu}(-t)) \nonumber \\
&&\hspace{4cm} -(\ha_{\nu}(-t)  \rho_{A}   \hx(-i \hbar \lambda-t) \hx   -  \rho_{A} \hx(-i \hbar \lambda-t) \ha_{\nu}(-t) \hx) \}, \nonumber \\
\label{eqn:13n} 
\end{eqnarray} 
respectively.  Here, we assumed that \(\langle \hy(t)\rangle=0\), and we used Eq.~(\ref{eqn:a14}), and definitions \(\rho_{A} \equiv \frac{1}{Z_{A}} e^{- \beta \ch_{A}}\) with \(Z_{A}= {\rm Tr}_{A} e^{- \beta \ch_{A}}\) and
\begin{eqnarray}
\Phi(t) \equiv {\rm Tr}_{R} \; \rho_{R}\; \hy(t)\; \hy \equiv \langle\, \hy(t)\; \hy \,\rangle.
\label{eqn:14n} 
\end{eqnarray} 

It might be convenient to use the eigenstates of an unperturbed relevant system to obtain the matrix elements of \(\rho_{A}(t)\) in the ``master" equation, Eq.~(\ref{eqn:7}).  We denote the eigenstates of the relevant system \(| l \rangle \) and \(| m \rangle \) for the energy eigenvalues as \(E_{l}\) and \(E_{m}\).  The \((l,m)\) component of \( \Xi_{2}(t-\tau) {\rho_{A}}_{\nu}(\tau) \) is given by
\begin{eqnarray}
\langle l |  \Xi_{2}(t-\tau) {\rho_{A}}_{\nu}(\tau) | m \rangle \nonumber \\
&& \hspace{-4cm}=- \sum_{k,n} 
[ \Phi(t-\tau)  \hx_{l,k} \hx(-(t-\tau))_{k,n} \{e^{-i \el_{S} (t-\tau)} {\rho_{A}}_{\nu}(\tau)\}_{n,m} \nonumber \\
&&\hspace{-1.5cm}-\Phi(t-\tau) \hx_{l,k} \{e^{-i \el_{S} (t-\tau)} {\rho_{A}}_{\nu}(\tau)\}_{k,n} \hx(-(t-\tau))_{n,m}\nonumber \\
&&\hspace{-1.5cm}-\Phi(-(t-\tau)) \hx(-(t-\tau))_{l,k} \{e^{-i \el_{S} (t-\tau)} {\rho_{A}}_{\nu}(\tau)\}_{k,n} \hx_{n,m} \nonumber \\
&&\hspace{-1.5cm}+\Phi(-(t-\tau)) \{e^{-i \el_{S} (t-\tau)} {\rho_{A}}_{\nu}(\tau)\}_{l,k} \hx(-(t-\tau))_{k,n} \hx_{n,m}]. \nonumber \\
\label{eqn:15n} 
\end{eqnarray} 
We can obtain the elements for \(\Psi_{2}(t)\) in a similar way.
In order to evaluate the susceptibility, Eq.~(\ref{eqn:10}), we need to obtain the Fourier-Laplace transform of each element and solve the simultaneous equations for Eq.~(\ref{eqn:7}). We can express the equation by making use of the Hilbert-Schmidt (or Liouville) space, which we show in the next subsection.

\subsection{Transformation to Hilbert-Schmidt space}
\label{sec:2c} 
In order to evaluate the susceptibility, Eq.~(\ref{eqn:10}), it is convenient to transform operators of the relevant system \(S\) into vectors that construct the H-S space. This is because the Liouville operators in the Hilbert space are written as a supermatrix in the H-S space, which makes the evaluations easier.  Defining a scalar product between operators \(\co\) and \(\cod_{i}\) as \( (\cod_{i},\co)={\rm Tr} \cod_{i}^{\dagger} \co \), the transformation from the Hilbert space to the H-S space is done by expanding an arbitrary operator \(\co\) in the Hilbert space with a set of orthonormal operators \({\cod_{i}}\) as
\begin{equation}
\co=\sum_{i} \cod_{i} (\cod_{i}, \co),
\label{eqn:16n}
\end{equation} 
where the orthonormal condition of \({\cod_{i}}\) is written as \((\cod_{i}, \cod_{j})=\delta_{i,j}\).   We can transform an operator in the Hilbert space to a vector in the H-S space with the set of \( (\cod_{i}, \co) \).  In the case where an operator \(\co\) is  written as an \(N\) dimensional matrix, the corresponding vector in the H-S space has \(N^2\) elements.  

Multiplication operations on a density operator in the Hilbert space are transformed to a supermatrix in the H-S space:  When the arbitrary operators \(\co_{1}\) and \(\co_{2}^{\dagger}\) are multiplied by the density operator \(\rho\) as \(\co_{1} \rho \co_{2}^{\dagger}\), the product is transformed into the H-S space as
\begin{equation}
\co_{1} \rho \co_{2}^{\dagger} \rightarrow \chm \; {\vec \rho},
\label{eqn:20n}
\end{equation} 
where the supermatrix \(\chm\) is symbolically expressed as 
\begin{equation}
\chm = \cm_{1} \otimes \cm_{2}^{*}. 
\label{eqn:21n}
\end{equation} 
Here \(\cm_{1}\) and \(\cm_{2}\) correspond to matrices of the operators \(\co_{1}\) and \(\co_{2}\), \(\otimes\) denotes the Kornecker product, and \(*\) denotes the complex conjugate operation. When the density operator is written in an \( N \times N \)-dimensional matrix, \(\rv(t)\) is an \( N^2 \)-dimensional vector and \(\chm_{\Xi_{2}}(t)\) is an \( N^2 \times N^2  \)-dimensional matrix.

Using Eqs.(\ref{eqn:20n}) and (\ref{eqn:21n}), we obtain the transformation of Eq.~(\ref{eqn:7}) into the H-S space as
\begin{equation}
\frac{d}{dt} \rv(t) = - \frac{i}{\hbar} \chm_{S} \rv(t) +
        \int_{t_{0}}^{t} d \tau \chm_{\Xi_{2}}(t-\tau) \rv(\tau)+ \pv_{2}(t),
\label{eqn:22n}
\end{equation}
which gives the susceptibility in a more straightforward way than using Eq.~(\ref{eqn:15n}), since the kernel, \(\Xi_{2}(t)\) is written in the H-S space as a matrix
\begin{eqnarray}
\chm_{\Xi_{2}}(t)&=&
-\Phi(t) [ \{\hx \hx(-t) e^{-\ih \ch_{S} t}\} \otimes (e^{-\ih \ch_{S} t})^{*} -\{ \hx e^{-\ih \ch_{S} t}\} \otimes \{\hx(-t)^{\dagger} e^{-\ih \ch_{S} t}\}^{*}]\nonumber \\
&&+\Phi(-t)[ \{\hx(-t) e^{-\ih \ch_{S} t}\} \otimes \{ \hx^{\dagger}  e^{-\ih \ch_{S} t}\}^{*} -e^{-\ih \ch_{S} t} \otimes \{\hx^{\dagger} \hx(-t)^{\dagger} e^{-\ih \ch_{S} t} \}^{*} ]. \nonumber \\
\label{eqn:23n} 
\end{eqnarray}

The inhomogeneous term in the H-S space is given by
\begin{eqnarray}
\pv_{2}(t)&=&i \int_{0}^{\beta} d\lambda \Phi(-i \hbar \lambda-t) \nonumber \\
&& \times \{
 \{\hx \ha_{\nu}(-t)\} \otimes \{\hx(i \hbar \lambda-t)\}^{*} - \{\hx\} \otimes \{\ha_{\nu}(-t) \hx(i \hbar \lambda-t)\}^{*}\nonumber \\
&&-( \{\ha_{\nu}(-t)\}  \otimes \{\hx  \hx(i \hbar \lambda-t)\}^{*} -  \v1 \otimes \{\hx \ha_{\nu}(-t) \hx(i \hbar \lambda-t)\}^{*}  ) \}\rva, 
\label{eqn:23n1} 
\end{eqnarray}
which is an \( N^2 \)-dimensional vector for the \( N \times N \)-dimensional density operator.

The  \(i(n,m)\)-th component of Eq.~(\ref{eqn:22n}) is expressed by
\begin{eqnarray}
\frac{d}{dt} \rv(t)_{i(n,m)} &=& -(\frac{i}{\hbar} \chm_{S} \rv(t))_{i(n,m)} \nonumber \\
&+& \int_{0}^{t} d \tau \chm_{\Xi_{2}}(t-\tau)_{i(n,m),j(n',m')} \rv(\tau)_{j(n',m')}+ \pv_{2}(t)_{i(n,m)},
\label{eqn:24n}
\end{eqnarray}
where \(i(n,m)=(n-1)N+m\) with \(n,m=1,2, \dots N\).
It should be noted that the Fourier-Laplace transform of the memory kernel in Eq.~(\ref{eqn:24n}) is given by
\begin{equation}
  \int_{0}^{\infty} dt e^{-i \omega t} \int_{0}^{t} d \tau \chm_{\Xi_{2}}(t-\tau)_{i(n,m),j(n',m')} \rv(\tau)_{j(n',m')} = \chm_{\Xi_{2}}[\omega]_{i(n,m),j(n',m')} \rv[\omega]_{j(n',m')},
\label{eqn:25n}
\end{equation}
and, we have
\begin{equation}
i \omega \rv[\omega] - \rv(0) = - \frac{i}{\hbar} \chm_{S} \rv[\omega] +\chm_{\Xi_{2}}[\omega] \rv[\omega] + \pv_{2}[\omega].
\label{eqn:26n} 
\end{equation} 
Thus, we have
\begin{equation}
\rv[\omega]= \chm_{\chi} (\rv(0)+\pv_{2}[\omega]),
\label{eqn:27n} 
\end{equation} 
with 
\begin{equation}
\chm_{\chi}=[i \omega +\frac{i}{\hbar} \chm_{S} -\chm_{\Xi_{2}}[\omega]]^{-1},
\label{eqn:28n} 
\end{equation} 
which corresponds to \((i \omega +  i \el_{S} - \Xi_{2}[ \omega])^{-1}\) in Eq.~(\ref{eqn:10}).  All of the matrix elements of \(\chm_{\chi}\) are given in an explicit way, as will be shown below.  The complex susceptibility in the H-S space is given in the form
\begin{equation}
\chi_{\mu \nu}(\omega)= \frac{i}{\hbar} ({\vec \hb}_{\mu} ,  \chm_{\chi} (\rv(0) + \pv_{2}[\omega])).
\label{eqn:29n} 
\end{equation} 

\subsection{Concrete form of \(\chm_{\Xi_{2}}[\omega]\)}
\label{sec:2d} 
Now we obtain the matrix elements of \(\chm_{\Xi_{2}}[\omega]\) on the basis of the eigenstates of the relevant system,
\begin{eqnarray}
\chm_{\Xi_{2}}[\omega]_{i(n,m),j(n',m')}\nonumber \\ 
&&\hspace{-3cm} =-\int_{0}^{\infty} dt e^{-i \omega t}  \Phi(t) [\sum_{k=1}^{N} X_{n,k} X_{k,n'} e^{-i (\omega_{k}-\omega_{m}) t} \delta_{m,m'} + X_{n,n'} e^{-i (\omega_{n'}-\omega_{m}) t} X^{*}_{m,m'}] \nonumber \\
&&\hspace{-3cm} +\int_{0}^{\infty} dt e^{-i \omega t}  \Phi(-t)[X_{n,n'} e^{-i (\omega_{n}-\omega_{m'}) t} X^{*}_{m,m'}+\sum_{k=1}^{N} X^{*}_{m,k} X^{*}_{k,m'} e^{-i (\omega_{n}-\omega_{k}) t} \delta_{n,n'}],
\label{eqn:30n} 
\end{eqnarray} 
with the eigenfrequency \(\omega_{l} \equiv E_{l}/\hbar\).  A more explicit expression is obtained by introducing the spectrum of the thermal bath, as
\begin{equation}
J(\omega)= \int_{-\infty}^{\infty} dt e^{-i \omega t} \Phi(t).
\label{eqn:31n} 
\end{equation} 
Using the spectrum,
\begin{eqnarray}
\chm_{\Xi_{2}}[\omega]_{i(n,m),j(n',m')} \nonumber \\ 
&&\hspace{-3cm} =- \frac{1}{2\pi} \int_{-\infty}^{\infty} d\omega'\{ [\sum_{k=1}^{N} X_{n,k} X_{k,n'} \delta_{m,m'} \kappa_{p}(\omega+\omega'+\omega_{km}) \nonumber \\ 
&&\hspace{0.5cm}+ X_{n,n'} X^{*}_{m,m'} \kappa_{p}(\omega+\omega'+\omega_{n'm}) ] J(\omega')  \nonumber \\
&& - [\sum_{k=1}^{N} X^{*}_{m,k} X^{*}_{k,m'} \delta_{n,n'} \kappa_{p}(\omega+\omega'+\omega_{nk})\nonumber \\
&&\hspace{0.5cm}+ X_{n,n'} X^{*}_{m,m'} \kappa_{p}(\omega+\omega'+\omega_{nm'})] J(-\omega')  \},
\label{eqn:32n} 
\end{eqnarray} 
where we define \(\omega_{km}=\omega_{k}-\omega_{m}\) and use the following relation, 
\begin{equation}
\lim_{\varepsilon \rightarrow +0}  \int_{0}^{\infty} dt \; e^{-i \omega t-\varepsilon t} = \pi \delta(\omega) - i \wp \frac{1}{\omega} \equiv \kappa_{p}(\omega).
\label{eqn:33n} 
\end{equation}
The terms of the principal value represent the frequency shift that results from the system-bath interaction.  While these terms have often been  neglected, we can take them into account in the present formalism.

\subsection{Concrete form of \(\pv_{2}[\omega]\)}
\label{sec:2e} 
The inhomogeneous term, Eq.~(\ref{eqn:13n}), is simply written in the H-S space by the multiplication of a matrix and the H-S vector of \(\rho_{A}\) 
\begin{equation}
\pv_{2}(t) \equiv \chm_{\pv_{2}}(t) \rva,
\label{eqn:33n1} 
\end{equation}
where 
\begin{eqnarray}
\chm_{\pv_{2}}(t)_{i(n,m),j(n',m')} \nonumber \\
&&\hspace{-4cm} = i \int_{0}^{\beta} d\lambda \Phi(-i \hbar \lambda-t) \{ 
\sum_{k}^{N} X_{n,k} {A_{\nu}}_{k,n'} X^{*}_{m,m'} e^{-i\omega_{kn'} t} e^{i(t+i \hbar \lambda ) \omega_{mm'}}  \nonumber \\
&& -X_{n,n'} \sum_{k}^{N} {A_{\nu}^{*}}_{m,k} {X^{*}}_{k,m'} 
e^{i \omega_{mk} t} e^{i(t+i \hbar \lambda ) \omega_{km'}} \nonumber \\
&&-( {A_{\nu}}_{n,n'} e^{-i \omega_{nn'} t} \sum_{k}^{N} X^{*}_{m,k} X^{*}_{k,m'} e^{i(t+i \hbar \lambda ) \omega_{km'}} \nonumber \\
&&-  \delta_{n,n'} \sum_{k,l}^{N} e^{i \omega_{kl}t} X^{*}_{m,k} {A_{\nu}^{*}}_{k,l} e^{i(t+i \hbar \lambda ) \omega_{lm'}} X^{*}_{l,m'}) \}. 
\label{eqn:23n1} 
\end{eqnarray}
The Fourier-Laplace transform of the inhomogeneous term \(\chm_{\pv_{2}}(t)\) is given as
\begin{eqnarray}
\chm_{\pv_{2}}[\omega]_{i(n,m),j(n',m')} &\equiv& \frac{i}{2\pi} \int_{-\infty}^{\infty} d\omega'  \{ 
\sum_{k}^{N} X_{n,k} {A_{\nu}}_{k,n'} X^{*}_{m,m'} \kappa_{p}(\omega'+\omega_{kn'}-\omega_{mm'}) \kappa_{i}(\omega'+\omega_{mm'})  \nonumber \\
&& -X_{n,n'} \sum_{k}^{N} {A_{\nu}^{*}}_{m,k} {X^{*}}_{k,m'} \kappa_{p}(\omega'-\omega_{mm'}) \kappa_{i}(\omega'+\omega_{km'}) \nonumber \\
&&-( {A_{\nu}}_{n,n'}  \sum_{k}^{N} X^{*}_{m,k} X^{*}_{k,m'} \kappa_{p}(\omega'+\omega_{nn'}-\omega_{km'}) \kappa_{i}(\omega'+\omega_{km'})\nonumber \\
&&-  \delta_{n,n'} \sum_{k,l}^{N}  X^{*}_{m,k} {A_{\nu}^{*}}_{k,l}  X^{*}_{l,m'} \kappa_{p}(\omega'-\omega_{kl}-\omega_{lm'}) \kappa_{i}(\omega'+\omega_{lm'})) \}, 
\label{eqn:33n} 
\end{eqnarray}
where we define \(\kappa_{i}(\omega)\) as
\begin{equation}
\int_{0}^{\beta} d\lambda e^{- \hbar \lambda \omega }  = \frac{1-e^{- \hbar \beta \omega }}{\omega} \equiv \kappa_{i}(\omega).
\label{eqn:33n} 
\end{equation}
The complex susceptibility, Eq.~(\ref{eqn:29n}), is now written in an explicit form, which can be applied to an arbitrary type of thermal bath by specifying \(J(\omega)\). 
We will show a few examples of baths in the next subsection.

\subsection{Bath}
\label{sec:2e} 
When analyzing the relaxation phenomena of a relevant system, we often introduce a thermal bath that consists of an infinite number of bosons\cite{leggett,weiss,leggett2} or spins\cite{prokofev}. This section discusses procedures to obtain the spectra  \(J(\omega)\) for a bosonic bath .
We use a bosonic bath for the relaxation phenomena caused by phonons in a medium or photons in a cavity. The Hamiltonian for the boson system is written as,
\begin{equation}
\chb=\sum_{\alpha} \hbar \omega_{\alpha} b_{\alpha}^{\dagger} b_{\alpha},
\label{eqn:34n} 
\end{equation} 
where \(b_{\alpha}\) (\(b^{\dagger}_{\alpha}\)) denotes an annihilation (creation) operator for the\(\alpha\)-th mode of a boson. 
As an example, we will consider a case in which the contribution to \(\ch_{1}\), Eq.~(\ref{eqn:11n}), from the bath is given by,
\begin{equation}
\hy \equiv \sum_{\alpha} g_{\alpha} (b_{\alpha}^{\dagger}+b_{\alpha}). 
\label{eqn:35n}
\end{equation} 
Then, the correlation function \(\langle\, \hy(t)\; \hy \,\rangle\) for the bath is written as
\begin{equation}
\Phi(t)=\sum_{\alpha}g_{\alpha}^2 \{\langle b_{\alpha} b^{\dagger}_{\alpha}\rangle e^{-i \omega_{\alpha} t} + \langle b^{\dagger}_{\alpha} b_{\alpha}\rangle e^{i \omega_{\alpha} t} \}, 
\label{eqn:36n} 
\end{equation} 
where \(g_{\alpha}\) is the coupling constant between the relevant system and the \(\alpha\)-th mode of the boson.  In order to evaluate the correlation function of the thermal bath, we need to introduce a coupling spectral function \(I(\omega)\) as 
\begin{equation}
I(\omega)=\sum_{\alpha}g_{\alpha}^2 \delta(\omega-\omega_{\alpha}).
\label{eqn:37n} 
\end{equation}
We can rewrite the weighted summation for an arbitrary function \(f(\omega_{\alpha})\) in the following form,
\begin{equation}
\sum_{\alpha}g_{\alpha}^2 f(\omega_{\alpha})=\int_{0}^{\infty } d\omega \sum_{\alpha}g_{\alpha}^2 \delta(\omega-\omega_{\alpha}) f(\omega) = \int_{0}^{\infty } d\omega I(\omega) f(\omega).
\label{eqn:37n} 
\end{equation}
Using Eq.~(\ref{eqn:37n}), \(\Phi(t)\) is rewritten as 
\begin{equation}
\Phi(t)=\int_{0}^{\infty} d\omega' I(\omega') \{ (n(\omega')+1) e^{-i \omega' t} + n(\omega') e^{i \omega' t} \},
\label{eqn:38n} 
\end{equation}
where \(n(\omega)\) is the boson distribution function given by \(n(\omega) = \frac{1}{e^{\beta \hbar \omega}-1}\). 
The spectrum of the thermal bath is obtained in the form
\begin{equation}
J(\omega)= \int_{-\infty}^{\infty} dt e^{i \omega t} \Phi(t) 
= I(\omega) [n(\omega)+1] \theta(\omega) + I(\omega) n(\omega) \theta(-\omega),
\label{eqn:39n} 
\end{equation} 
where \(\theta(\omega)\) denotes the step function.

\section{Applications}
\label{sec:4} 
We are now in a position to apply the formalism presented in the previous section to the relaxation phenomena in a spin system.  First, we evaluate the spectral line shape of a system where a single spin interacts with a bosonic bath. Although such a system is trivial, the evaluation shows the concrete procedure, which is essentially the same as in multiple-spin systems.  Next, we demonstrate the Nagata-Tazuke effect for two and three spin systems, showing the dependence of the line shape on the angle between the spatial configuration and the direction of the static applied field.  We include the initial correlation and the frequency shift of the line shapes. 

\subsection{Spin-boson model}
\label{sec:4a} 
Suppose that a spin (\(S=\frac{1}{2}\)) linearly interacts with a thermal bath that consists of bosons. The Hamiltonian of the relevant system is written as
\begin{equation}
\cha= \hbar \omega_{0} S_{z},
\label{eqn:43n} 
\end{equation} 
and the interaction operator \(\hx\) in Eq.~(\ref{eqn:11n}) is given by
\begin{equation}
\hx \equiv a S_{x} + c S_{z} 
\label{eqn:44n} 
\end{equation} 
where \(S_{m}, (m=x,y,z) \) corresponds to the \(x,y\), and \(z\) components of the relevant spin. In the following, we set \(a= \sin \Lambda\)  and \(c=\cos \Lambda\), since the generality is not lost when \(a\) is a real number.   We control the types of relaxation by the value of \(\Lambda\):  the case of \(\Lambda=0\) \((a=0, c=1)\) describes the pure dephasing phenomena of the spin.  For other cases of \(\Lambda \ne 0\), we can include the longitudinal relaxation in the transverse relaxation of the spin.

Equation~(\ref{eqn:28n}) is now given as follows:  The second term is written as  
\begin{eqnarray}
\frac{i}{\hbar}\chm_{S}&=& \frac{i}{\hbar}(\ch_{S} \otimes \v1 - \v1 \otimes \ch_{S}^{*}) = \left(
\begin{array}{llll}
 0 & 0 & 0 & 0 \\
 0 & i \omega_{0} & 0 & 0 \\
 0 & 0 & -i \omega_{0} & 0 \\
 0 & 0 & 0 & 0
\end{array}
\right).  \label{eqn:45n} 
\end{eqnarray}
We can evaluate the third term of Eq.~(\ref{eqn:28n}) concretely by using Eq.~(\ref{eqn:32n}) as, 

\begin{eqnarray}
\chm_{\Xi_{2}}[\omega] \nonumber \\
&&{\hspace{-2cm}} =- \frac{1}{4}
\left(
\begin{array}{llll}
 |a|^2  \phi_{1}[\omega,\omega_{0}]  & -a c ( 2 Fs_{-}[\omega,\omega_{0}]+\phi_{3}[\omega,0] ) & - a^{*} c (2 F_{+}[\omega,\omega_{0}]-\phi_{3}[\omega,0] )   &  - |a|^2 \phi_{1}[\omega,\omega_{0}]  \\
-  a^{*} c {\phi_{4}}_{-}[\omega,\omega_{0}]  & |a|^2 {\phi_{4}}_{+}[\omega,0]+ 2 c^2 {\phi_{4}}_{-}[\omega,\omega_{0}] &
- {a^{*}}^2 {\phi_{4}}_{+}[\omega,0] &  a^{*} c {\phi_{4}}_{-}[\omega,\omega_{0}] \\   
-a c {\phi_{4}}_{+}[\omega,\omega_{0}]  & -  a^2 {\phi_{4}}_{+}[\omega,0] &  |a|^2 {\phi_{4}}_{+}[\omega,0] + 2 c^2 {\phi_{4}}_{+}[\omega,\omega_{0}] & a c {\phi_{4}}_{+}[\omega,\omega_{0}] \\
-|a|^2 \phi_{2}[\omega,\omega_{0}]) & a c (2 F_{-}[\omega,\omega_{0}] - \phi_{3}[\omega,0]) & a^{*} c (2 Fs_{+}[\omega,\omega_{0}]+ \phi_{3}[\omega,0]) & |a|^2 \phi_{2}[\omega,\omega_{0}]
\end{array}
\right),\nonumber \\
\label{eqn:46n} 
\end{eqnarray} 
where
\begin{eqnarray}
\phi_{1}[\omega,\omega_{0}] &\equiv& F_{+}[\omega,\omega_{0}] + Fs_{-}[\omega,\omega_{0}]\;\;\;,\;\;\;
\phi_{2}[\omega,\omega_{0}] \equiv Fs_{+}[\omega,\omega_{0}] + F_{-}[\omega,\omega_{0}], \nonumber \\
\phi_{3}[\omega,\omega_{0}]&\equiv& F_{+}[\omega,\omega_{0}] - Fs_{+}[\omega,\omega_{0}]\;\;\;,\;\;\;
{\phi_{4}}_{\pm}[\omega,\omega_{0}] \equiv F_{\pm}[\omega,\omega_{0}] +Fs_{\pm}[\omega,\omega_{0}]
\label{eqn:47n} 
\end{eqnarray} 
and
\begin{eqnarray}
F_{\pm}[\omega,\omega_{0}] &\equiv & \int_{0}^{\infty} dt \Phi(t) e^{i (\pm \omega_{0}-\omega)t} \nonumber \\ &=& \pi \{I(-\omega \pm \omega_{0}) (n(-\omega \pm \omega_{0})+1) \theta (-\omega \pm \omega_{0}) \nonumber \\
&&\hspace{0.5cm}+ I(-(-\omega \pm \omega_{0}))n(-(-\omega \pm \omega_{0})) \theta(-(-\omega \pm \omega_{0}))\}\nonumber \\
&&-i \wp \int_{0}^{\infty} d\omega' \left(\frac{1}{\omega \mp \omega_{0}+\omega'} I(\omega') (n(\omega')+1) + \frac{1}{\omega \mp \omega_{0}-\omega'} I(\omega') n(\omega')\}\right), \nonumber \\ \label{eqn:48n} \\ 
Fs_{\pm}[\omega,\omega_{0}] &\equiv & \int_{0}^{\infty} dt \Phi^{*}(t) e^{i (\pm \omega_{0}-\omega)t} \nonumber \\ &=& \pi \{I(\omega \mp \omega_{0}) (n(\omega \mp \omega_{0})+1) \theta (\omega \mp \omega_{0}) \nonumber \\
&&\hspace{0.5cm}+ I(-(\omega \mp \omega_{0}))n(-(\omega \mp \omega_{0})) \theta(-(\omega \mp \omega_{0}))\}\nonumber \\
&&-i \wp \int_{0}^{\infty} d\omega' \left(\frac{1}{\omega \mp \omega_{0}-\omega'} I(\omega') (n(\omega')+1) + \frac{1}{\omega \mp \omega_{0}+\omega'} I(\omega') n(\omega')\}\right). \nonumber \\
\label{eqn:49n} 
\end{eqnarray} 
It should be noted that the principal value  integrals are  included in Eq.~(\ref{eqn:49n}).  
The inhomogeneous term is given by
\begin{eqnarray}
\pv_{2}[\omega] &=& 
2 i \times \left(
\begin{array}{l}
a^{*} c {A_{\nu}}_{+}  \eta_{1-}[\omega] +  a c {A_{\nu}}_{-} \eta_{1+}[\omega]  - |a|^2 {A_{\nu}}_{z}  \left( \eta_{2-}[\omega] + \eta_{2+}[\omega] \right)\\
 2 a^{*} c {A_{\nu}}_{z}  \eta_{2+}[\omega] + {a^{*}}^2 {A_{\nu}}_{+} \eta_{3-}(t)- {A_{\nu}}_{-} \left(2 c^2 \eta_{1+}[\omega] + |a|^2 \eta_{3+}(t)\right)\\
2 a c{A_{\nu}}_{z}  \eta_{2-}(t)+a^2 {A_{\nu}}_{-}  \eta_{3+}(t)
    - {A_{\nu}}_{+} \left(2 c^2 \eta_{1-}[\omega] + |a|^2 \eta_{3-}(t)\right)\\
- a c {A_{\nu}}_{-} \eta_{1+}[\omega] - a^{*} c {A_{\nu}}_{+} \eta_{1-}[\omega] + |a|^2 {A_{\nu}}_{z} \left( \eta_{2-}[\omega] + \eta_{2+}[\omega] \right) 
\end{array}
\right), \label{eqn:50n} 
\end{eqnarray} 
where \(\eta_{\mu,\pm}[\omega] (\mu=1,2,3)\) are given in Appendix B.

Next, we show a numerical evaluation of the susceptibility for the Ohmic coupling spectral function  
\begin{equation}
I(\omega)=  s \; \omega \; e^{-\omega/\omega_{c}},
\label{eqn:51n} 
\end{equation} 
where \(s\) denotes the coupling strength and \(\omega_{c}\) denotes the cut-off frequency.  

The concrete techniques used in the numerical evaluation of the formula,  Eq.~(\ref{eqn:29n}), are as follows.  First, we avoided the matrix inversion  procedure in Eq.~(\ref{eqn:28n}). By rewriting the complex susceptibility in
the form,
\begin{equation}
\chi_{\mu \nu}(\omega)= \frac{i}{\hbar} ({\vec \hb}_{\mu} ,  {\vec x})
\end{equation} 
with
\begin{equation}
 {\vec x} \equiv \chm_{\chi} (\rv(0) + \pv_{2}[\omega]),
\end{equation} 
we find that the essential task required in Eq.~(\ref{eqn:29n}) is obtaining the vector \({\vec x}\)
by solving the simultaneous equation for the elements
\begin{equation}
[i \omega + \frac{i}{\hbar} \chm_{S} -\chm_{\Xi_{2}}[\omega]] {\vec x}=(\rv(0) + \pv_{2}[\omega]).
\end{equation}
Second, in solving the above equation, we numerically calculated the principal value integral in Eqs.~(\ref{eqn:48n}) and (\ref{eqn:49n}) by using a Mathematica built-in function.   Because this evaluation should be performed carefully, we checked the results by comparing them with those obtained by the trapezoidal numerical integration method.   In the present form of \(I(\omega)\), Eq.~(\ref{eqn:51n}), we have an analytically evaluated correlation function  \(\Phi(t)\),
\begin{equation}
\Phi(t)=\frac{s \omega_{c}^2 (1-\omega_{c}^2 t^2)}{(1+\omega_{c}^2 t^2)^2}+\frac{2 s}{\hbar^2 \beta^2} \{ \psi' (1+\frac{1}{\hbar \beta \omega_{c}}+\frac{i t}{\hbar \beta})+\psi' (1+\frac{1}{\hbar \beta \omega_{c}}-\frac{i t}{\hbar \beta})\}-\frac{2 i s \omega_{c}^3 t}{(1+\omega_{c}^2 t^2)^2}.
\label{eqn:61na} 
\end{equation} 
The function \(\psi' (z)=\frac{d}{dz} \psi(z)\) in Eq.~(\ref{eqn:61na}) is defined by the digamma function, \(\psi(z)= \frac{\Gamma'(z)}{\Gamma(z)} \) as in \cite{Allahverdyan}. 
Using this form, we also checked the above mentioned numerical estimations of the time integrals in Eqs.~(\ref{eqn:48n}) and (\ref{eqn:49n}). We confirmed that all three of the estimations gave the same result.

\begin{figure}[h]
\begin{center}
\includegraphics[scale=1.0]{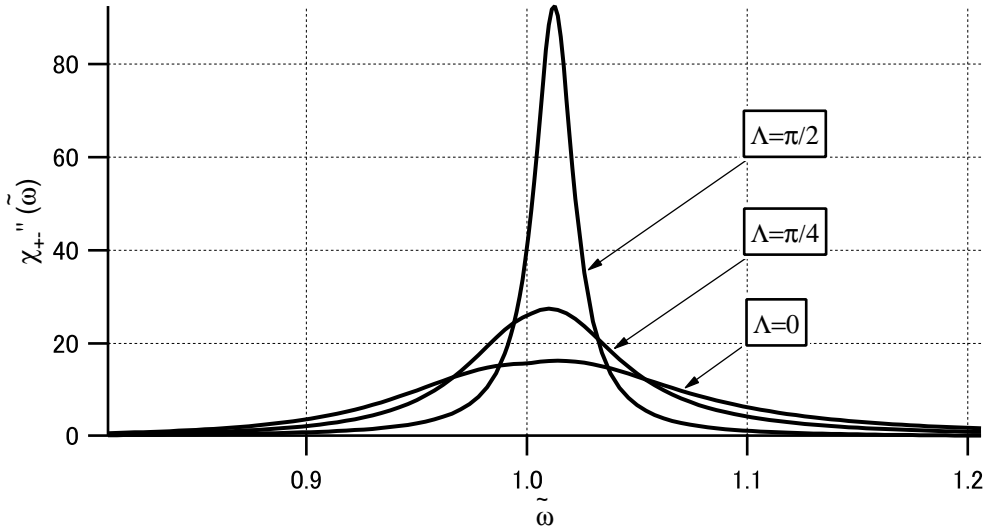}
\end{center}
\caption{Transverse susceptibility \(\chi_{+-}''(\omega)\) for \(k_{B} T= \hbar \omega_{0}/5\), \(s=0.1\) and \({\tilde \omega}_{c}=0.5\) with changing \(\Lambda\) as \(0,\frac{\pi}{4}, \frac{\pi}{2}\).}
\label{fig:fig1}
\end{figure}

Figure \ref{fig:fig1} shows the imaginary part of the transverse susceptibility \(\chi_{+-}''(\tw)\), found by using Eq.~(\ref{eqn:29n}) as a function of the frequency of the external field scaled by the Larmor frequency of the spin \(\omega_{0}\),  i.e., \(\tw \equiv \omega/\omega_{0}\). We scaled the coupling strength \(s\) and the cut-off frequency \(\omega_{c}\)  with \(\omega_{0}\), and set them to be \(s=0.1\) and \({\tilde \omega}_{c} \equiv \omega_{c} /\omega_{0}=0.5\).   We also set the temperature of the bath to be \(k_{B} T= \hbar \omega_{0}/5\).  
We study the three cases of \(\Lambda=0,\frac{\pi}{4}, \frac{\pi}{2}\), which determine the types of relaxation:  the case of \(\Lambda=0\) corresponds to the adiabatic interaction case (i.e., the pure dephasing case),  \(\Lambda=\frac{\pi}{4}\) to the transverse relaxation case, and \(\Lambda=\frac{\pi}{2}\) to the non-adiabatic interaction case. 

Figure \ref{fig:fig1} shows that the width of the spectra decreases with increasing \(\Lambda\).  This is explained as follows: for \(\Lambda=0\), the thermal bath affects the spin as a random magnetic field along the \(z-\)axis.  The direction of this random magnetic field tilts toward the \(x-\)axis as \(\Lambda\) increases to \(\frac{\pi}{2}\). 

We can explain this fact by remembering the relaxation time obtained in the Markovian limit, where the transverse relaxation time \(T_{2}\) is given by\cite{us93,hss},
\begin{equation}
\frac{1}{T_{2}}=\frac{1}{2} \left( \frac{1}{T_{1}} + \frac{1}{\tau_{0}} \right).
\label{eqn:52n} 
\end{equation} 
Equation~(\ref{eqn:52n}) was obtained for the system-bath interaction as
\begin{equation}
\ch_{1}=\hbar g {\vec S} \cdot {\vec R},
\label{eqn:55n} 
\end{equation} 
where \({\vec S}\) and \({\vec R}\) are the relevant spin and bath operator, respectively.  The relaxation times \(T_{1}\) and \(\tau_{0}\) are given by
\begin{equation}
\frac{1}{T_{1}}=2 {\rm Re} (\varphi_{+-}+\varphi_{-+}^{*}), \;\;\; \frac{1}{\tau_{0}}=2 {\rm Re} \, \varphi_{zz},
\label{eqn:53n} 
\end{equation} 
where 
\begin{equation}
\varphi_{\pm \mp}=\frac{g^2}{4}\int_{0}^{\infty} dt e^{\mp i \omega_{0} t} \langle R_{\pm}(t) R_{\mp}(0) \rangle, \;\;\;
\varphi_{z z}= g^2 \int_{0}^{\infty} dt \langle R_{z}(t) R_{z}(0) \rangle, 
\label{eqn:54n} 
\end{equation}
with \( R_{\pm} =R_{x} \pm i R_{y}\).
Comparing Eqs.~(\ref{eqn:11n}) and (\ref{eqn:44n}) with Eq.~(\ref{eqn:55n}), we find that the case of \(\Lambda=0\) corresponds to \(g {\vec R}=(0,0,g R_{z})=(0,0,\hy)\) and \(\Lambda=\frac{\pi}{2}\) to \(g {\vec R}=(g R_{x},0,0)=(\hy,0,0)\). When we consider an extreme case of \(\langle R_{x}(t) R_{x}(0) \rangle =\langle R_{z}(t) R_{z}(0) \rangle \varpropto \delta(\omega_{c} t)\), we find that \(\frac{1}{T_{2}}=\frac{1}{2 \tau_{0}}=\frac{g^2}{2\omega_{c}}\) for \(\Lambda=0\), and \(\frac{1}{T_{2}}=\frac{1}{2 T_{1}}=\frac{g^2}{4\omega_{c}}\) for \(\Lambda=\frac{\pi}{2}\).  This means that the width of the transverse spectrum decreases with increasing \(\Lambda\), which essentially explains the physical origin of the fact shown in Fig.\ref{fig:fig1}.

It should be noted that the type of system-bath interaction in Eq.~(\ref{eqn:55n}) is different from the one in Eq.~(\ref{eqn:11n}), except for cases where the thermal-bath operator \({\vec R}\) is described as \({\vec R}={\hat R} {\vec n}\) with an arbitrary vector \({\vec n}\).  Since the vector is written as \({\vec n}={\vec z}({\vec x})\) in the case of \(\Lambda=0\) (\(\Lambda=\frac{\pi}{2}\)), we can consider the above correspondence. While we can easily extend the system-bath interaction to \( \ch_{1} = \sum_{i} \hx_{i} \, \hy_{i}\), as in Eq.~(\ref{eqn:55n}), we chose the simple form of Eq.~(\ref{eqn:11n}) for a demonstration.

\begin{figure}[h]
\begin{center}
\includegraphics[scale=1.0]{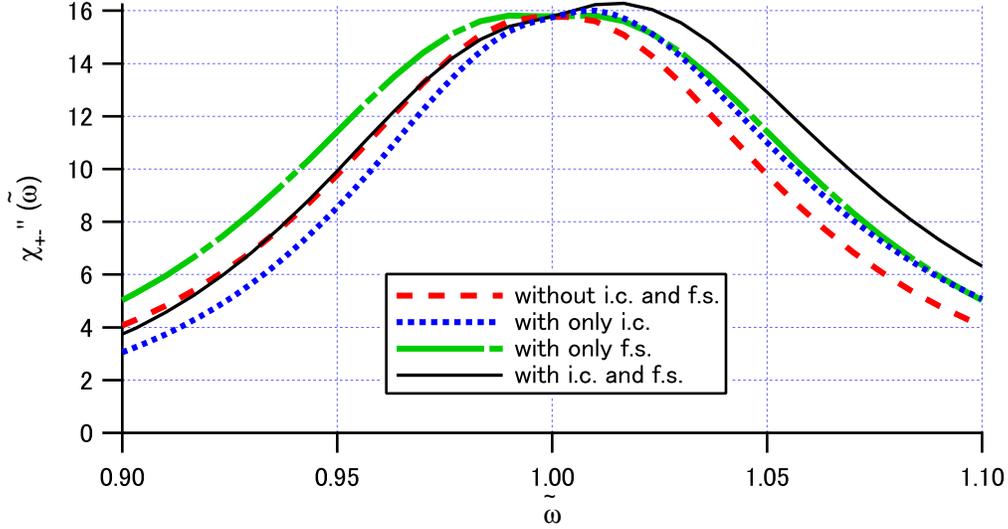}
\end{center}
\caption{(Color Online)The effects of the initial correlation and frequency shift on the transverse susceptibility \(\chi_{+-}''(\omega)\) for \(\Lambda=0\).   The other parameters are the same as in Fig.\ref{fig:fig1}.  The solid (black) line refers to the evaluation with the initial correlation and frequency shift, the dot-dash (green) line refers to the evaluation with just the frequency shift, the dotted (blue) line refers to the evaluation with just the initial correlation, and the broken (red) line refers to the evaluation without the initial correlation and frequency shift.  In this figure, we abbreviate the initial correlation as i.c. and the frequency shift as f.s.}
\label{fig:fig2}
\end{figure}

In Fig.\ref{fig:fig1}, we can see the higher frequency shift due to the imaginary part of \( \Xi_{2}[ \omega] \). The detailed structures of the shifts can be seen by comparing the results with and without the effects of the initial correlation and the frequency shift in Figs.\ref{fig:fig2} \(\sim\) \ref{fig:fig4}. Since the broken (red) lines in these figures have peaks at \(\tw=1\), we find that the initial correlation and frequency shift cause the spectra to shift to the higher frequencies.  The spectral shift and shape depend on the coupling strength \({\tilde s}\), and a detailed analysis of this dependence will be presented in a forthcoming paper.

\begin{figure}[h]
\begin{center}
\includegraphics[scale=1.0]{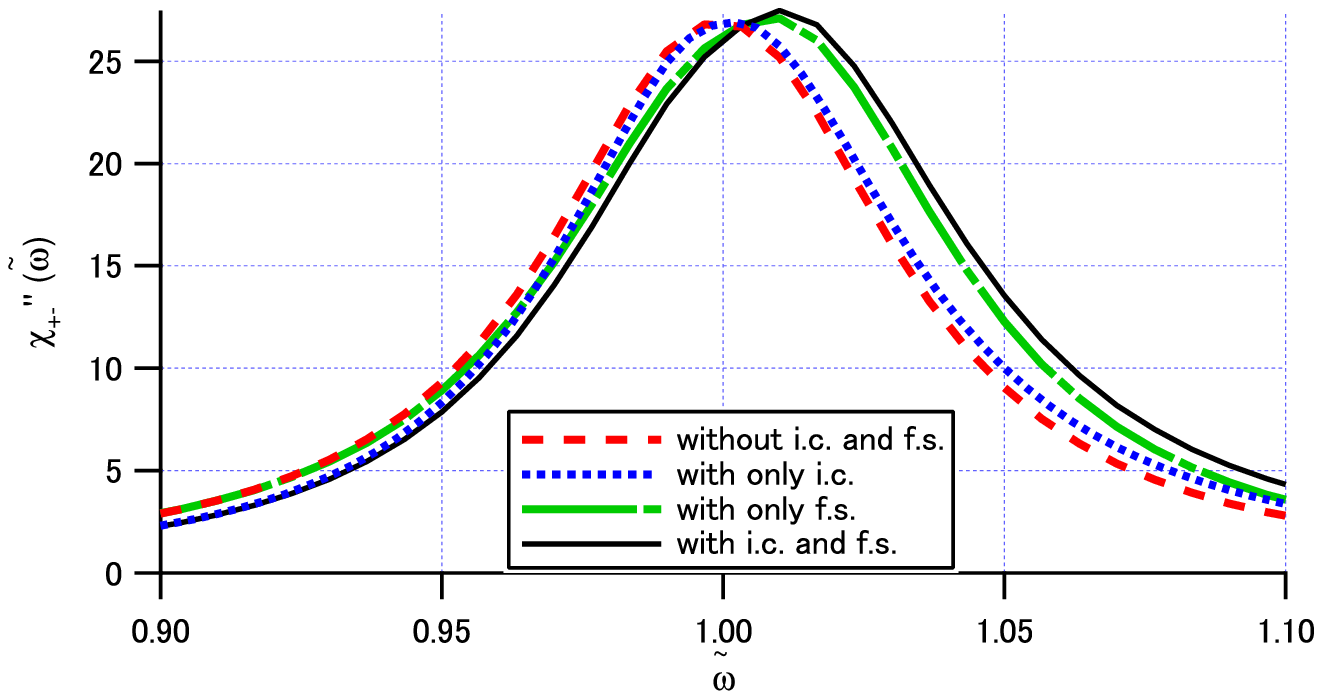}
\end{center}
\caption{(Color Online)The effects of the initial correlation and frequency shift on the transverse susceptibility \(\chi_{+-}''(\omega)\) for \(\Lambda=\frac{\pi}{4}\)).  The parameters and notations of lines are the same as in Fig.\ref{fig:fig2}.}
\label{fig:fig3}
\end{figure}
\begin{figure}[h]
\begin{center}
\includegraphics[scale=1.0]{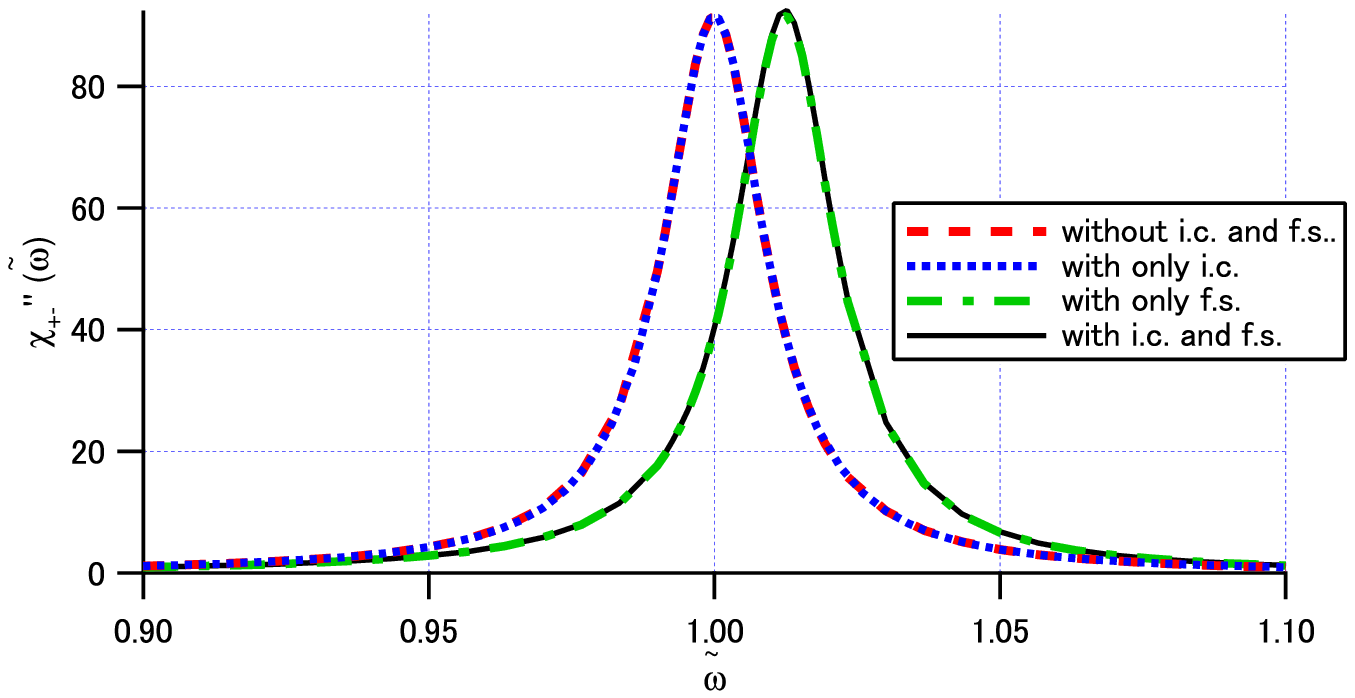}
\end{center}
\caption{(Color Online)The effects of the initial correlation and frequency shift on the transverse susceptibility \(\chi_{+-}''(\omega)\) for \(\Lambda=\frac{\pi}{2}\)).  The parameters and notations of lines are the same as in Fig.\ref{fig:fig2}.}
\label{fig:fig4}
\end{figure}

\subsection{Interacting spin-boson model}
\label{sec:3b} 
Next, we discuss the absorption spectra of interacting \(N\) spins in contact with a bosonic bath, focusing our attention on the types of spin systems that break the
Heisenberg SU(2) symmetry. These interactions cause shifts from the paramagnetic resonance spectra, called resonance shifts\cite{nagata}.  Typical examples of such interactions are the anisotropic exchange interaction and the dipole-dipole interaction. 
In these cases, the Hamiltonian of the relevant system is as follows, 
\begin{equation}
\cha=\hbar \omega_{0} \sum_{i}^{N} S_{i,z} +\ch_{ex}+\ch_{D}, \;\;\label{eqn:63n}
\end{equation}
where \(S_{i,z}\) denotes the \(z\) component of the \(i\)th spin and \(\ch_{ex}\) is the exchange interaction written by
\begin{equation}
\ch_{ex}=-2 \hbar J \sum_{i,j} (S_{i,x}S_{j,x}+S_{i,y}S_{j,y}+A S_{i,z}S_{j,z}),\;\;\label{eqn:64n}
\end{equation}
with exchange interaction energy \(J\), and anisotropy parameter \(A\).
\(\ch_{D}\) in Eq.~(\ref{eqn:63n}) is the dipole-dipole interaction given by
\begin{equation}
\ch_{D}=D \sum_{i,j} \frac{1}{r_{ij}^3} \left\{\si \cdot \sj- \frac{3}{r_{ij}^2}(\si \cdot \rij)(\sj \cdot \rij) \right\},\;\;\label{eqn:65n} \\
\end{equation}
where \(\rij\) is the vector from the spin \(i\) to the spin \(j\), \(r_{ij}=|\rij|\).  In Eq.~(\ref{eqn:65n}), we define \(D= \frac{\mu_{0}}{4 \pi} (\mu_{s})^2\) where \(\mu_{0}\) is the magnetic permeability and \(\mu_{s}\) is the magnitude of the magnetic moment that carries the relevant spin.  When we consider an electron(nuclear)-spin, we have \(\mu_{s}=g_{e} \mu_{B}(g_{n} \mu_{n})\), where \(g_{e}(g_{n})\) is the \(g\)-value of an electron (nuclear) and \(\mu_{B} (\mu_{n})\) is the Bohr magneton (nuclear magneton). 

We can investigate the resonance shift due to these interactions by considering how the spins interact linearly with a bosonic bath, as
\begin{eqnarray}
\chab&=& \hbar \hx \hy,  \label{eqn:66n}\\
\hx &\equiv&  \sum_{i}^{N} \frac{1}{2} (a_{i}^{*} S_{i,+} + a_{i} S_{i,-}) + c_{i} S_{i,z}, \label{eqn:67n} \\
\hy &\equiv& \sum_{\alpha} g_{\alpha} (b_{\alpha}^{\dagger}+b_{\alpha}), \label{eqn:68n}
\end{eqnarray} 
where we define \(a_{i}= e^{i \Lambda_{2,i}} \sin \Lambda_{1,i} \) and \(c_{i}=\cos \Lambda_{1,i}\), which control the interactions between the \(i\)-th spin and the bath.
\subsubsection{Two-spin system}
\begin{figure}[h]
\begin{center}
\includegraphics[scale=0.6]{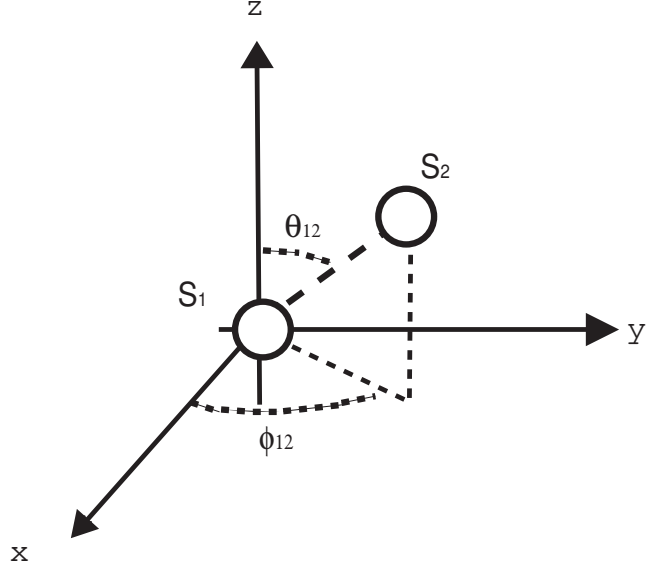}
\end{center}
\caption{Angles in spherical coordinates for two spins \(S_{1}\) and \(S_{2}\).}
\label{fig:fig5}
\end{figure}

When the relevant system consists of two spins, the spin-spin interaction portion of Eq.~(\ref{eqn:63n}) can be rewritten as
\begin{equation}
\ch_{ex}+\ch_{D}= \hbar \left( {\begin{array}{*{20}c}
   {S_{1,x} \,,} & {S_{1,y} \,,} & {S_{1,z} } \end{array}\,} \right)
\left( {\begin{array}{*{20}c}
   {h_{11} } & {h_{12} } & {h_{13} }  \\
   {h_{21} } & {h_{22} } & {h_{23} }  \\
   {h_{31} } & {h_{32} } & {h_{33} }  \\
\end{array}} \right) 
\left( {\begin{array}{*{20}c}
   {S_{2,x} }  \\
   {S_{2,y} }  \\
   {S_{2,z} }  \\
\end{array}} \right),
 \label{eqn:69n}
\end{equation}
where
\begin{eqnarray}
h_{ii}&\equiv &-2(J+D_{0}(\Omega_{i}^2-1/3)) ,\;\;\; (i =1,2) \nonumber \\
h_{33}&\equiv & -2(A J+D_{0}(\Omega_{3}^2-1/3)) \nonumber \\
h_{ij}&\equiv & h_{ji}=-2 D_{0} \Omega_{i} \Omega_{j}, \;\;\; (i \ne j),\
 \label{eqn:70n}
\end{eqnarray}
with \(D_{0}\equiv \frac{3 D}{2 r_{12}^3 \hbar}\).  Here we define 
\begin{equation}
\frac{\r12}{r_{12}}=\left( {\begin{array}{*{20}c}
   \Omega_{1}   \\
   \Omega_{2}   \\
   \Omega_{3}   \\
\end{array}} \right) =\left( {\begin{array}{*{20}c}
   {\sin \theta _{12} \,\cos \phi _{12} }  \\
   {\sin \theta _{12} \,\sin \phi _{12} }  \\
   {\cos \theta _{12} }  \\
\end{array}} \right),
 \label{eqn:71n}
\end{equation}
where \(\theta_{12}\) and \(\phi _{12}\) are the angles of  \(\r12\) in spherical coordinates(see Fig.\ref{fig:fig5}).  

\begin{figure}[h]
\begin{center}
\includegraphics[scale=1.0]{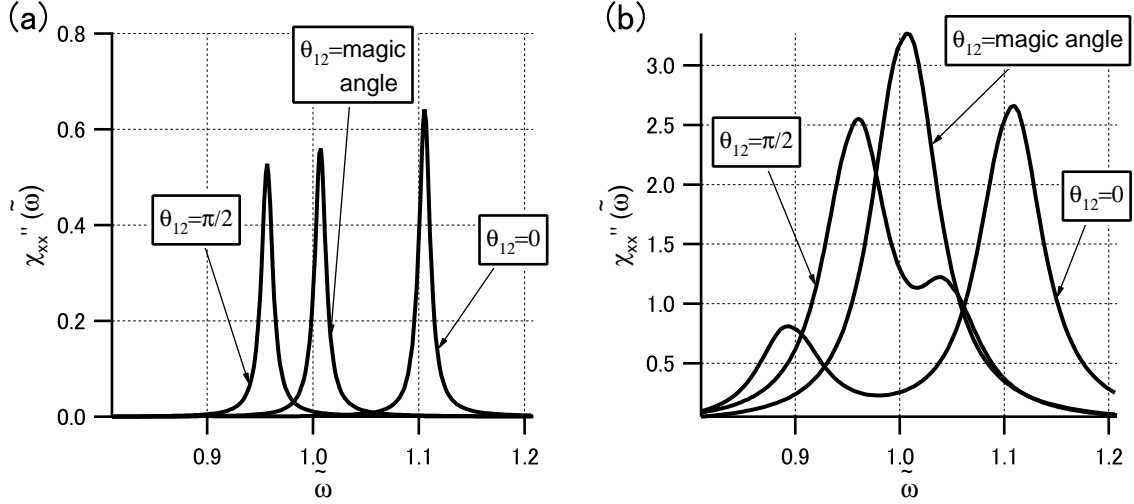}
\end{center}
\caption{Transverse susceptibility \(\chi_{xx}''(\omega)\) by changing \(\theta _{12}\) from \(0\) to \(\frac{\pi}{2}\) with \(\phi _{12}=0\).  The other parameters were set as \({\tilde \omega}_{c}=0.5\), \(s=1/50\), \({\tilde D_{0}}=0.1\), \({\tilde J}=-1\), and \(A=1.0\). (a) shows the lower temperature case, \(k_{B}T= \hbar \omega_{0}/5\) and (b) shows the higher temperature case, \(k_{B}T= \hbar \omega_{0}\).}
\label{fig:fig6}
\end{figure}

We can show the dependence of the line shape on the angle \(\theta _{12}\) when keeping \(\phi_{12}=0\).  In Fig.\ref{fig:fig6}, we show this effect for the isotropic exchange interaction, \(A=1.0\), and for the weak dipole-dipole interaction, which is scaled by the Larmor frequency of the relevant spin as, \({\tilde D_{0}}\equiv \frac{D_{0}}{\omega_{0}}=0.1\). The exchange interaction, which is scaled by the Larmor frequency, is set to be \({\tilde J}\equiv \frac{J}{\omega_{0}}=-1\). We set the scaled cut-off frequency as \({\tilde \omega}_{c}=0.5\), and the coupling strength as \(s=0.02\).  We consider the type of spin relaxation to be pure dephasing by setting \(\Lambda_{1,i}=\Lambda_{2,i}=0 \) with \(i=1,2\).   Figure \ref{fig:fig6}(a) shows the case of a lower temperature \(k_{B}T= \hbar \omega_{0}/5\), where we can see a sharp peak, which shows a lower frequency shift as \(\theta _{12}\) increases from \(0\) to \(\frac{\pi}{2}\) via the magic angle \((=\arccos(\frac{1}{\sqrt{3}}))\). For a higher temperature, \(k_{B}T= \hbar \omega_{0}\), we find that an additional peak appears for \(\theta _{12}=0\) and \(\frac{\pi}{2}\) to give asymmetric spectra in Fig.\ref{fig:fig6}(b). In the evaluations for Fig.\ref{fig:fig6}, we include the effects of both the initial correlation and frequency shift, and find the peak shifts as in the lower temperature case. 

We can explain this peak shift behavior using quantum mechanical evaluations.  As typical examples, let us take two cases of \((\theta_{12}, \phi_{12})=(0,0)\) and \((\pi/2,0)\), for which the interaction Hamiltonian Eq.~(\ref{eqn:69n}) becomes diagonal.  Defining the elements of \(\{h_{ii}\}\) with \(i=1 \sim 3\) as \((h_{11},h_{22},h_{33})=-2 (J_{x}^{{\rm eff}}, J_{y}^{{\rm eff}}, J_{z}^{{\rm eff}})\) respectively, we have for \((\theta_{12}, \phi_{12})=(0,0)\), 
\begin{equation}
J_{x}^{{\rm eff}}=J_{y}^{{\rm eff}}=J-D_{0}/3 \,\,,\,\,J_{z}^{{\rm eff}}=J+2 D_{0}/3,
\label{eqn:72n}
\end{equation}
and for \((\theta_{12}, \phi_{12})=(\pi/2,0)\) ,
\begin{equation}
J_{x}^{{\rm eff}}=J+2 D_{0}/3\,\,,\,\,J_{y}^{{\rm eff}}=J_{z}^{{\rm eff}}=J-D_{0}/3.
\label{eqn:73n}
\end{equation}
Using the eigenvectors of \(S_{z}\), \(|\pm\rangle\), which correspond to the eigenvalues \(\pm \hbar \omega_{0}/2\), we can obtain the eigenvalues and eigenvectors of the system Hamiltonian, Eq.~(\ref{eqn:63n}), in the form,
\begin{eqnarray}
E_{a}&=&\hbar (-j_{z}+K)  \,\,,\,\,\,\,\,\,\,\,\,\,|a\rangle= \frac{1}{\sqrt{2 K (K+\omega_{0})}} [(j_{x}-j_{y}) |++\rangle - (K+\omega_{0}) |--\rangle], \nonumber \\
E_{b}&=&\hbar (j_{z}-j_{x}-j_{y}) \,\,,\,\,\,|b\rangle=\frac{1}{\sqrt{2}} [|+-\rangle + |-+\rangle], \nonumber \\
E_{c}&=&\hbar(-j_{z}-K) \,\,,\,\,\,\,\,\,\,\,\,\,|c\rangle=\frac{1}{\sqrt{2 K (K+\omega_{0})}} [(j_{x}-j_{y}) |++\rangle + (K+\omega_{0}) |++\rangle], \nonumber \\
E_{d}&=&\hbar(j_{z}+j_{x}+j_{y}) \,\,,\,\,\,|d\rangle=\frac{1}{\sqrt{2}} [|+-\rangle - |-+\rangle], \nonumber \\
\label{eqn:74n}
\end{eqnarray}
where we denote \(j_{\mu}=J_{\mu}^{{\rm eff}}/2\) with \(\mu=x,y,z\), and \(K=\sqrt{(\omega_{0})^2+(j_{x} -j_{y})^2}\).  

\begin{figure}[h]
\begin{center}
\includegraphics[scale=1.0]{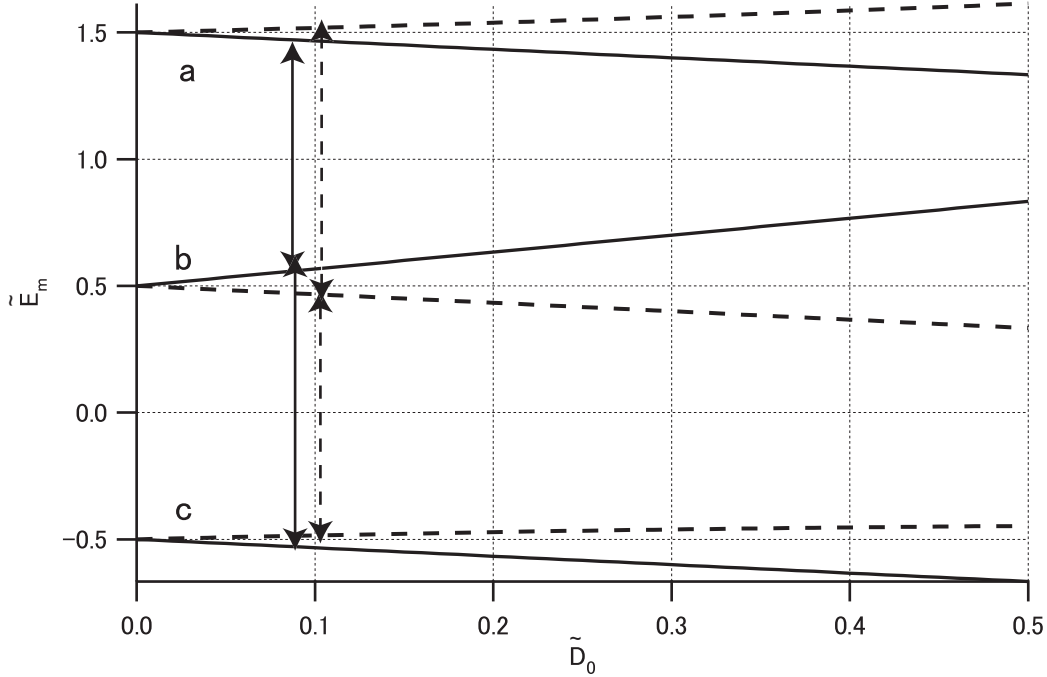}
\end{center}
\caption{The dependence of the eigenvalues \({\tilde E}_{m}(=E_{m}/\hbar)\) with \(m=a \sim c\) on the scaled dipole-dipole interaction \({\tilde D_{0}}\) for \(\theta_{12}=0\)(solid line) and \(\theta_{12}=\frac{\pi}{2}\)(dashed line). The other parameters are the same as in Fig.\ref{fig:fig6}. }
\label{fig:fig7}
\end{figure}

In Fig.\ref{fig:fig7}, we show the dependence of the eigenvalues \({\tilde E}_{m}(\equiv E_{m}/\hbar)\) with \(m=a,b,c\) on the scaled dipole-dipole interaction \({\tilde D_{0}}\) for \(\theta_{12}=0\) and \(\theta_{12}=\frac{\pi}{2}\). The other parameters are the same as in Fig.\ref{fig:fig6}.  The solid (dashed) lines refer to the energy eigenvalues for  \(\theta_{12}=0\) (\(\theta_{12}=\frac{\pi}{2}\)). Comparing the eigenstates of the isotropic Heisenberg system, which are obtained in the limit of \(D_{0} \rightarrow 0\), we can consider that the states \(|a\rangle, |b\rangle\) and \(|c\rangle\) correspond to the triplet states, \(|1,-1\rangle, |1,0\rangle\) and \(|1,1\rangle\), respectively. (The state \(|d\rangle\) corresponds to the singlet state, \(|0,0\rangle\).)   

The peaks in  Fig.\ref{fig:fig6} reflect the transitions between the triplet states for \({\tilde D_{0}}=0.1\):  the peaks in Fig.\ref{fig:fig6}(a) correspond to the transition between \(|b\rangle\) and \(|c\rangle\).  The solid (dashed) arrows in  Fig.\ref{fig:fig7} refer to the energy differences for \(\theta_{12}=0\) (\(\theta_{12}=\frac{\pi}{2}\)) around \({\tilde D_{0}}=0.1\). Due to the fact that the \({\tilde E}_{b}\) and \({\tilde E}_{c}\) for  \(\theta_{12}=\frac{\pi}{2}\) bend inside more than for \(\theta_{12}=0\), the length of the dashed arrow is longer than the length of the solid arrow.  This means that the peak frequency for \(\theta_{12}=\frac{\pi}{2}\) is lower than that for \(\theta_{12}=0\) in Fig.\ref{fig:fig6}(a). We also have the other type of transition, between the triplet states, \(|a\rangle\) and \(|b\rangle\).  However, the transition probability is very small in the lower temperature range, as in Fig.\ref{fig:fig6}(a).  The amplitude becomes larger with an increase in temperature, resulting in additional peaks, which correspond to the transition between the triplet states, \(|a\rangle\) and \(|b\rangle\) in Fig.\ref{fig:fig6}(b).

The peak shift of ESR due to the dipole-dipole interaction in one-dimensional antiferromagnets was theoretically explained by Nagata and Tazuke\cite{nagata}. They investigated the absorption spectra by changing the direction of the magnetic field \(H_{0}\) from \(H_{0}//c\) to \(H_{0}\bot c\).    
In experiments,  the line shape is usually given as a function of \(H_{0}\), but not the frequency \(\omega\).  Therefore, we have to evaluate the dependence of the line shape based on the strength of the static magnetic field, rather than the frequency, as in Fig.\ref{fig:fig6}.  The correspondence is discussed in the Appendix C.  

\subsubsection{Three-spin system}
\begin{figure}[h]
\begin{center}
\includegraphics[scale=0.6]{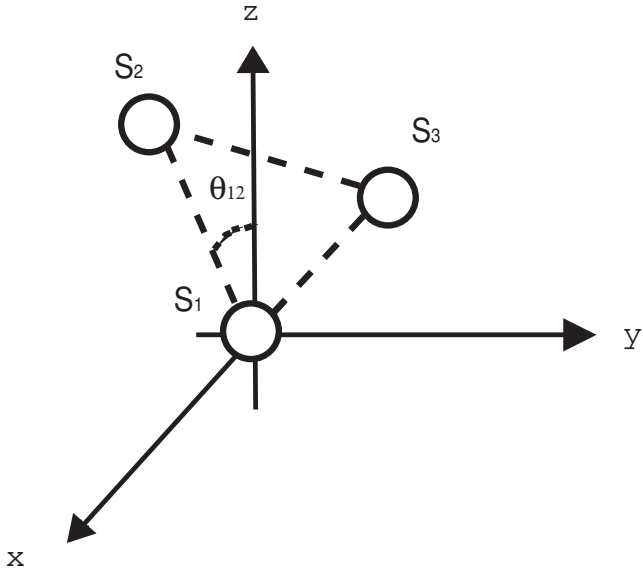}
\end{center}
\caption{Arrangement of 3 spins drawn as spheres.}
\label{fig:fig8}
\end{figure}

We now discuss the effect of dipole-dipole interaction in a relevant system with three spins (\(S=\frac{1}{2}\)) that form an equilateral triangle.
This is an extension of the Nagata-Tazuke shift to a triangle system. We can find a typical example in the antiferromagnetic triangular spin rings of \({\rm Cu}\) \cite{Nojiri}.
In order to study the  peak shift for these three spins, we incline the face of the triangle from the \(yz\)-plane to the \(xy\)-plane by increasing the angle \(\theta_{12}\) from \(0\) to \(\frac{\pi}{2}\) with \(\phi_{12}=0\) and keeping the normal of the triangle parallel to the \(x\)-axis, as shown in Fig.\ref{fig:fig8} .   

\begin{figure}[h]
\begin{center}
\includegraphics[scale=1.0]{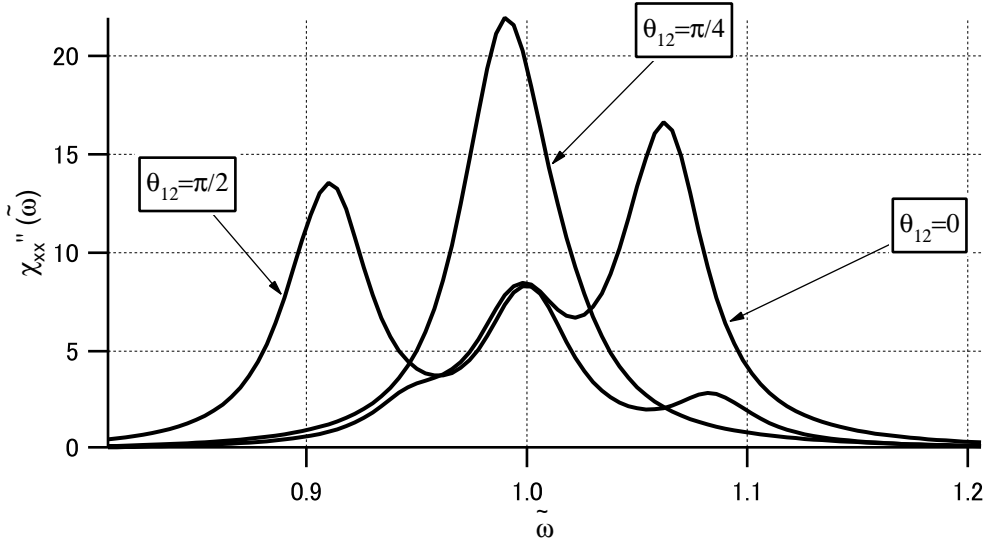}
\end{center}
\caption{Transverse susceptibility \(\chi_{xx}''(\omega)\) when changing \(\theta _{12}\) from \(0\) to \(\frac{\pi}{2}\) with \(\phi _{12}=0\).  The other parameters are set as \({\tilde \omega}_{c}=0.5\), \(s=1/150\), \({\tilde D_{0}}=0.1\), \({\tilde J}=1\),\(A=1.0\), and \(k_{B}T=\hbar \omega_{0}\).}
\label{fig:fig9}
\end{figure}

Figure \ref{fig:fig9} shows the line shapes of transverse susceptibility, which include the effects of the initial correlation and frequency shift by system-bath interaction.  Here we set the exchange interaction \({\tilde J} =1\),  the dipole-dipole interaction \({\tilde D_{0}}=0.1\), the isotropic exchange interaction \(A=1.0\), and \(k_{B}T=\omega_{0}\). We consider the spin relaxation type to be the pure dephasing by setting \(\Lambda_{1,i}=\Lambda_{2,i}=0 \) with \(i=1,2\).  We find three peaks in the cases where \(\theta _{12}=0\) and \(\theta _{12}=\frac{\pi}{2}\).  As \(\theta _{12}\) increases, the higher peak shifts from right to left.

\begin{figure}[h]
\begin{center}
\includegraphics[scale=1.0]{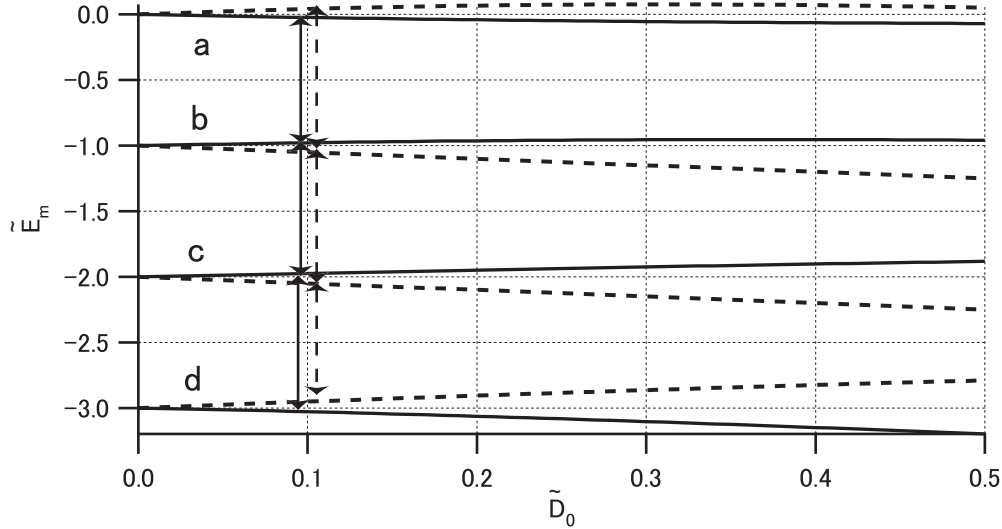}
\end{center}
\caption{The dependence of the eigenvalues of the quartet-levels (\({\tilde E}_{m}(=E_{m}/\hbar)\) with \(m=a \sim d\) on the scaled dipole-dipole interaction \({\tilde D_{0}}\). The other parameters are the same as in Fig.\ref{fig:fig9}. The solid line represents \(\theta_{12}=0\) and the dashed line represents \(\theta_{12}=\frac{\pi}{2}\). }
\label{fig:fig10}
\end{figure}

We can explain the peak shift behavior based on the angle dependence of the energy levels of the relevant system.  These levels consist of the lower four levels in the quartet states and the higher four levels in the doublet states, which are almost degenerate.  Figure \ref{fig:fig10} shows the dependence of the lower quartet-levels (\(a \sim d\)) on the scaled dipole-dipole interaction \({\tilde D_{0}}\) for \(\theta_{12}=0\) and \(\theta_{12}=\frac{\pi}{2}\). The other parameters are the same as in Fig.\ref{fig:fig9}. The solid lines (arrows) refer to the energy (differences) for  \(\theta_{12}=0\), and the dashed lines (arrows) refer to the energy (differences) for  \(\theta_{12}=\frac{\pi}{2}\).  These arrows are placed around \({\tilde D_{0}}=0.1\), which corresponds to the case in Fig.\ref{fig:fig9}. The length of the solid arrow between \(|c\rangle\) and \(|d\rangle\) is longer than the dashed arrow between \(|c\rangle\) and \(|d\rangle\). Since the transition probability between the lower energy levels becomes higher for the relatively lower temperature, we can consider that the highest peaks in Fig.\ref{fig:fig9} for \(\theta_{12}=0\) and \(\theta_{12}=\frac{\pi}{2}\) correspond to the transition between \(|c\rangle\) and \(|d\rangle\) in the quartet states. The fact that the length of the dashed arrow between \(|c\rangle\) and \(|d\rangle\) is shorter than the length of the solid arrow between \(|c\rangle\) and \(|d\rangle\) shows the reason for the shift in the highest peaks in Fig.\ref{fig:fig9}. Similarly, the lengths of the arrows between \(|a\rangle\) and \(|b\rangle\) show the shifts of the lowest peaks for \(\theta_{12}=0\) and \(\theta_{12}=\frac{\pi}{2}\).   

\section{Discussion and concluding remarks}
\label{sec:5} 
In this paper, we extend a linear response formula to include the frequency shift and initial correlation between the relevant system and the thermal bath.  Using the projection operator method, we show that the time evolution of the response function can be described with a Nakajima-Zwanzig type of equation.  We transform the equation into the Hilbert-Schmidt space to give a tractable formula where the density matrices are described with vectors, and the super operators are transformed into matrices.  The obtained formula enables us to systematically study the line shapes for various kinds of the system-bath interactions at arbitrary temperatures.  Moreover, with this formula it is easy to extend the relevant system to include multiple interacting spins.  We show the line shapes for a single and for two and three interacting spins, which suffer from the environmental effects of a bosonic bath.

The obtained formula enables us to evaluate the spectra, including the following three effects: (1) the non-Markovian effects of system-bath interaction, (2) the frequency shift by the system-bath interaction, and (3) the effects of the initial correlation between the relevant system and the thermal bath. 

While (2) and (3) are often neglected, the roles of these effects on the steady state of the reduced density operator have been studied up to the second order of interaction\cite{mori2008}.  It has been pointed out that these effects are essential to ensure the modification of the steady state by the system-bath interaction, and to prevent the steady state of the reduced density operator from approaching the equilibrium state of the relevant system.  Since we need the stationary response to the external oscillating field, it is necessary to include all of these effects in the time evolution of the response function to obtain the line shapes.

We apply the present formalism to spin systems interacting with a bosonic bath.  For a single spin system, we study the dependence of the line shape on the type of system-bath coupling, e.g., the pure dephasing case and the case of non-adiabatic interaction. We find that the initial correlation and frequency shift by the system-bath interaction are more significant in the pure dephasing case than in the non-adiabatic interaction case.  For two and three spin systems, we demonstrate the dependence of the line shape on the angle between the spatial spin arrangement and the direction of the static field.

We compare the obtained complex susceptibility in this paper with the conventional one in the Born-Markovian approximation in the Appendix D. Evaluating the transverse susceptibility for the spin-boson model in the pure dephasing case (a = 0 and c = 1 in Eq.~(\ref{eqn:44n})), we find that the frequency shift and the initial correlation cause the considerable peak shiftwhich are not included in the conventional Born-Markovian approximation.

Since the formula is written with the convolution integral as a Nakajima-Zwanzig type of master equation, we can systematically extend the formula to the higher orders of perturbation for the case of strong system-bath interaction\cite{uchiyama-shibata1,uchiyama-shibata2}.  One of the authors studied this problem in a strongly coupled spin-boson model, and it was pointed out that the time evolution of the transverse component of the spin was obtained in a closed form on the assumption of non-adiabatic system-bath interaction and a Lorentzian type of coupling spectral function, which enables us to systematically obtain the absorption spectra, including up to infinite orders of interaction.
In the present formula, a similar extension is possible. Moreover, the transformation into the Hilbert-Schmidt space enables us to obtain a form of absorption spectra on demand, even for the case of the interaction of multiple spin systems in any spatial arrangement.  We hope that our formalism will be a useful tool for analyzing the dynamics of various types of interacting spin systems.

\begin{acknowledgements}
This work was 
supported in part by Grant-in-Aid for Scientific Research on Priority Areas ``Physics of new quantum phases in superclean materials" 
(Grant No.\ 17071011),  Grant-in-Aid for Scientific Research (B) ``Analyses of nano-system with quantum statistical mechanical methods" (Grant No.\ 18340113), and by the Next Generation Super Computer Project, Nanoscience Program from MEXT.
\end{acknowledgements}

\appendix
\section{Derivation of Eq.~(\ref{eqn:7})}
\label{sec:a1}
We use the projection operator method to obtain the reduced dynamics of \({\rho_{A}}_{\nu}(t)\).  However, it should be noted that the reduced density operator \({\rho_{A}}_{\nu}(t)\) is different from the ordinary reduced density operator.  It is defined as \(\rho(t) \equiv {\rm Tr_{R}} W(t) ={\rm Tr_{R}} e^{-i \el t} W(0) \).  Here \(W(t)\) denotes the density operator of the whole system, which follows the Liouville von-Neuman equation,
\begin{equation}
\dw(t)=-i \el W(t).
\label{eqn:a1}
\end{equation} 
In this appendix, we show that we can obtain the ``master" equation for \({\rho_{A}}_{\nu}(t)\) by focusing on the time evolution operator \(e^{-i \el t}\) and extracting the relevant part from it. 

Let us define a projection operator \(\cp\), which eliminates the variables of the thermal bath, to obtain the relevant part of the time evolution operator.  The projection operator satisfies the idempotent relation, \(\cp^2=\cp\).  We also introduce a complementary operator \(\cq \equiv 1- \cp\).  Here we follow the standard method for deriving the equation of motion of the reduced operator\cite{Kubo,Nakajima,Zwanzig,uchiyama-shibata1}.  Denoting the relevant and irrelevant parts of the time evolution operator as 
\begin{equation}
x(t)\equiv  \cp e^{-i \el t}, \;\;\;  y(t)\equiv \cq e^{-i \el t},
\label{eqn:a2}
\end{equation} 
with an initial time \(t_{0}=0\), we obtain  
\begin{equation}
\frac{d}{dt} x(t)=\cp (- i \el) x(t) + \cp (- i \el) y(t) \; ,
\label{eqn:a3}
\end{equation}
and
\begin{equation}
\frac{d}{dt} y(t)=\cq (- i \el) x(t) + \cq (- i \el) y(t) \; .
\label{eqn:a4}
\end{equation}
Equation (\ref{eqn:a4}) has the following formal solution
\begin{equation}
y(t)=\int_{0}^{t} e^{-\cq i \el (t-\tau)} \cq (-i \el) x(\tau) d\tau + e^{-\cq i \el t} \cq.
\label{eqn:a5}
\end{equation}
Substituting Eq.~(\ref{eqn:a5}) into Eq.~(\ref{eqn:a3}), we obtain 
\begin{equation}
\frac{d}{dt} x(t)= \cp (-i \el) x(t)+\cp (-i \el)\int_{0}^{t} e^{-\cq i \el (t-\tau)} \cq (-i \el) x(\tau) d\tau+\cp (-i \el) e^{-\cq i \el t} \cq.
\label{eqn:a6}
\end{equation}

We set the specific form of the projection operator to be \(\cp=\rho_{{\rm R}} {\rm Tr_{R}}\), where \(\rho_{{\rm R}}\) denotes the density operator of the thermal bath, which is in the equilibrium state.  When we multiply Eq.~(\ref{eqn:a6}) by the initial density operator of the whole system \(W(t_{0})\), from the right hand side, we obtain the ordinary master equation for the reduced density operator \(\rho(t)\).  Instead of \(W(t_{0})\), we could also multiply Eq.~(\ref{eqn:a6}) by \([\ha_{\nu}, W_{eq}]\), which gives the ``master" equation for \({\rho_{A}}_{\nu}(t)\) in the form 
\begin{equation}
\frac{d}{dt} {\rho_{A}}_{\nu}(t) = - \frac{i}{\hbar} [\cha, {\rho_{A}}_{\nu}(t)] +
                           \int_{0}^{t} d \tau \xi (t-\tau) {\rho_{A}}_{\nu}(\tau)+\psi(t),
\label{eqn:a7}
\end{equation} 
where the kernel \(\xi(t)\) and the inhomogeneous term \(\psi(t)\) are given by
\begin{eqnarray}
\xi (t) &\equiv& \cp (-i \el) e^{-\cq i \el t} \cq (-i \el), 
\label{eqn:a8} \\
\psi(t) &\equiv& \cp (-i \el) e^{-\cq i \el t} \cq [\ha_{\nu}, W_{eq}].
\label{eqn:a9}
\end{eqnarray} 

Using the relations
\begin{equation}  
e^{-\cq i \el t}=e^{-\cq i \el_{0} t} T_{+}{\exp[\int_{0}^{t} dt' e^{i \el_{0} t'} \cq (-i \el_{1}) \cq e^{-i \el_{0} t'}]}
\label{eqn:a10} 
\end{equation} 
and \(\cp \el_{0} = \el_{0} \cp\),
the kernel \(\xi(t)\) in Eq.~(\ref{eqn:a8}) is written as
\begin{eqnarray} 
\xi (t) &=& \cp (-i \el) e^{- i \el_{0} t} \cq T_{+}{\exp[\int_{0}^{t} dt' e^{i \el_{0} t'} \cq (-i \el_{1}) \cq e^{-i \el_{0} t'}]} \cq (-i \el), 
\label{eqn:a11} 
\end{eqnarray} 
which gives the expansion of the kernel \(\xi(t)= \sum_{i=2} \Xi_{i} (t)\).  In   
Eq.~(\ref{eqn:a10}), \(T_{+}\) denotes the time ordering operator from right to left.
Taking up to the second order in \(\el_{1}\), we obtain 
\begin{eqnarray}
\int_{0}^{t} d\tau \Xi_{2} (t-\tau) {\rho_{A}}_{\nu}(\tau) 
&=& \int_{0}^{t} d\tau \cp (-i \el_{1}) e^{-i \el_{0} (t-\tau)} \cq (-i \el_{1}) {\rho_{A}}_{\nu}(\tau), 
\label{eqn:a12} 
\end{eqnarray} 
where we use the relation \(\cp \cq = \cq \cp=0 \).

The density operator of the whole system \(W_{eq}\) in \(\psi(t)\) is expanded as
\begin{equation}
W_{eq}=\frac{1}{Z} e^{-\beta (\ch_{0}+\ch_{1})}
=\frac{1}{Z} e^{-\beta \ch_{0}} (1-\int_{0}^{\beta} d\lambda \ch_{1}(-i \hbar \lambda) + \cdots),
\label{eqn:a13} 
\end{equation} 
where
\begin{equation}
\ch_{1}(t)= e^{\frac{i}{\hbar} \ch_{0} t} \ch_{1} e^{-\frac{i}{\hbar} \ch_{0} t}.
\label{eqn:a14} 
\end{equation} 
Using Eqs.~(\ref{eqn:a10}) and Eq.~(\ref{eqn:a13}), we obtain the expansion of \(\psi (t) \equiv \sum_{i=1} \Psi_{i}(t)\). The low order terms are given by
\begin{eqnarray}
\Psi_{1}(t)&=&0 , \label{eqn:a15} \\
\Psi_{2}(t)&=& \cp (-i \el_{1}) e^{-i \el_{0} t} \cq [\ha_{\nu}, - \rho_{0} \int_{0}^{\beta} d\lambda \ch_{1}(-i \hbar \lambda)], \label{eqn:a16}
\end{eqnarray}
where \(\rho_{0} \equiv \frac{1}{Z_{0}} e^{-\beta \ch_{0}}\) with \(Z_{0} \equiv {\rm Tr}_{S+B} e^{-\beta \ch_{0}}\).  To obtain Eq.~(\ref{eqn:a16}), we take up to the second order in \(\ch_{1}\) by expanding the partition function \(Z\) for the total system as
\begin{eqnarray}
Z&=& {\rm Tr}_{S+B} e^{-\beta (\ch_{0}+\ch_{1})} \nonumber \\
 &=& {\rm Tr}_{S+B} e^{-\beta \ch_{0}} (1-\int_{0}^{\beta} d\lambda \ch_{1}(-i \hbar \lambda) + \cdots) \nonumber \\
 &=&  Z_{0} + {\rm Tr}_{S+B} e^{-\beta \ch_{0}} (-\int_{0}^{\beta} d\lambda \ch_{1}(-i \hbar \lambda) + \cdots).
\label{eqn:a17} 
\end{eqnarray} 
Using Eqs.~(\ref{eqn:a7}), (\ref{eqn:a12}), and (\ref{eqn:a16}), we obtain the ``master" equation for \({\rho_{A}}_{\nu}(t)\) in the form of Eq.~(\ref{eqn:7}).       

\section{Functions of \(\eta_{\mu,\pm}[\omega] (\mu=1,2,3)\)}
\label{sec:b1}
The explicit forms of \(\eta_{\mu,\pm}[\omega], (\mu=1,2,3)\) in Eq.~(\ref{eqn:50n}) are given as follows.
\begin{eqnarray}
\eta_{1,\pm}[\omega] &=& \cosh{\frac{\beta  \hbar  \omega_{0}}{2}} \{ \frac{\pi (1-e^{-\beta \hbar (\omega \pm \omega_{0})})}{\hbar (\omega \pm \omega_{0})} \{I(\omega \pm \omega_{0}) (n(\omega \pm \omega_{0})+1) \theta (\omega \pm \omega_{0}) \nonumber \\
&&\hspace{5.5cm}+ I(-(\omega \pm \omega_{0}))n(-(\omega \pm \omega_{0})) \theta(-( \omega \pm \omega_{0}) ) \} \nonumber \\
&&-i \wp \int_{0}^{\infty} d\omega'\frac{1-e^{-\beta \hbar \omega'}}{\hbar \omega'}  \left(\frac{1}{\omega \mp \omega_{0}-\omega'} I(\omega') (n(\omega')+1) - \frac{1}{\omega \mp \omega_{0}+\omega'} I(\omega') n(\omega')\}\right)\}, \nonumber \\
\label{eqn:B1} \\
\eta_{2,\pm}[\omega] &=& e^{-\frac{\beta  \hbar  \omega_{0}}{2}} \{ \frac{\pi (1-e^{-\beta \hbar \omega})}{\hbar \omega} \{I(\omega \pm \omega_{0}) (n(\omega \pm \omega_{0})+1) \theta (\omega \pm \omega_{0}) \nonumber \\
&&\hspace{5.5cm}- I(-(\omega \pm \omega_{0}))n(-(\omega \pm \omega_{0})) \theta(-( \omega \pm \omega_{0}) ) \} \nonumber \\
&&-i \wp \int_{0}^{\infty} d\omega' ( \frac{1-e^{-\beta \hbar (\omega' \mp \omega_{0})}}{\hbar (\omega' \mp \omega_{0})} \frac{1}{\omega \mp \omega_{0}-\omega'} I(\omega') (n(\omega')+1)  \nonumber \\
&&\hspace{5.5cm}- \frac{1-e^{\beta \hbar (\omega' \pm \omega_{0})}}{\hbar (\omega' \pm \omega_{0})} 
\frac{1}{\omega \pm \omega_{0}+\omega'} I(\omega') n(\omega') ) ) \}, \nonumber \\
\label{eqn:B2} \\
\eta_{3,\pm}[\omega] &=& e^{\frac{\beta  \hbar  \omega_{0}}{2}} \{ \frac{\pi (1-e^{-\beta \hbar (\omega \pm \omega_{0})})}{\hbar (\omega \pm \omega_{0})} \{I(\omega) (n(\omega)+1) \theta (\omega) \nonumber \\
&&\hspace{5.5cm}- I(-\omega )n(-\omega) \theta(-\omega) \} \nonumber \\
&&-i \wp \int_{0}^{\infty} d\omega' ( \frac{1-e^{-\beta \hbar (\omega' \pm \omega_{0})}}{\hbar (\omega' \pm \omega_{0})} \frac{1}{\omega - \omega'} I(\omega') (n(\omega')+1)  \nonumber \\
&&\hspace{5.5cm}- \frac{1-e^{\beta \hbar (\omega' \mp \omega_{0})}}{\hbar (\omega' \mp \omega_{0})} 
\frac{1}{\omega + \omega'} I(\omega') n(\omega') ) ) \}, 
\label{eqn:B3} \\
\eta_{4,\pm}[\omega] &=& \{ \frac{\pi (1-e^{-\beta \hbar (\omega \pm 2 \omega_{0})})}{\hbar (\omega \pm 2 \omega_{0})} \{I(\omega \pm \omega_{0}) (n(\omega \pm \omega_{0})+1) \theta (\omega \pm \omega_{0}) \nonumber \\
&&\hspace{5.5cm}+ I(-(\omega \pm \omega_{0}))n(-(\omega \pm \omega_{0})) \theta(-( \omega \pm \omega_{0}) ) \} \nonumber \\
&&-i \wp \int_{0}^{\infty} d\omega' ( \frac{1-e^{-\beta \hbar (\omega' \pm \omega_{0})}}{\hbar (\omega' \pm \omega_{0})} \frac{1}{\omega - \omega' \pm \omega_{0}} I(\omega') (n(\omega')+1)  \nonumber \\
&&\hspace{5.5cm}- \frac{1-e^{\beta \hbar (\omega' \mp \omega_{0})}}{\hbar (\omega' \mp \omega_{0})} 
\frac{1}{\omega + \omega' \pm \omega_{0}} I(\omega') n(\omega') ) ) \}.
\label{eqn:B4} 
\end{eqnarray}
     
\section{Correspondence with the ESR experiments}
\label{sec:d}
For a two-spin system, we found that the peaks of the spectra moved to lower frequencies when  \(\theta _{12}\) increased from \(0\) to \(\frac{\pi}{2}\). In order to compare the ESR experiment for one-dimensional antiferromagnets by Nagata and Tazuke\cite{nagata}, we evaluated the spectra as a function of the magnitude of the static magnetic field \(H_{0}\) for a given frequency \(\omega\) of the oscillating field.  

\begin{figure}[h]
\begin{center}
\includegraphics[scale=0.8]{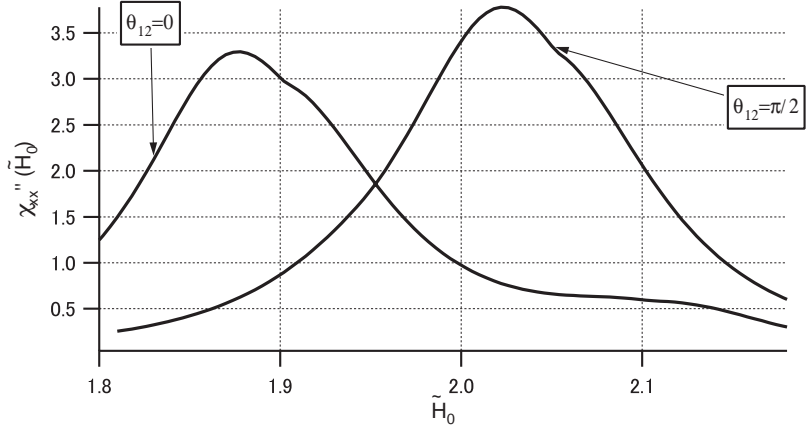}
\end{center}
\caption{Transverse susceptibility \(\chi_{xx}''(\omega)\) by changing \(\theta _{12}\) from \(0\) to \(\frac{\pi}{2}\) with \(\phi _{12}=0\).  The other parameters are set as \({\tilde J}' =-1\), \({\tilde \omega}_{c}'=0.5\), \(s=0.02\), \({\tilde D_{0}}'=0.1\), and \(A=1.0\).}
\label{fig:fig11}
\end{figure}

If a peak appears at \(\omega=\gamma H_{0} +\Delta \omega\) as a function of \(\omega\) (Fig.\ref{fig:fig6}(b)), i.e., 
\begin{equation}
\omega_{peak}(H_{0})=\gamma H_{0}+\Delta \omega,
\end{equation}
a peak in the shape of a function of \(\omega\)
\begin{equation}
H_{0}^{peak}=\frac{\omega_{peak}}{\gamma}-\frac{\Delta \omega}{\gamma},
\end{equation}
where \(\frac{\omega_{peak}}{\gamma}\), gives the position of the paramagnetic resonance.  Therefore, the peak moves in the opposite direction when we give the line shape as a function of \(H_{0}\). We give an example in Fig.\ref{fig:fig11}, where we adopted an oscillating field with a constant frequency \(\frac{\omega}{|J|}=2\). As the horizontal axis of the figure, we scaled the magnitude of the static magnetic field \(H_{0}\) with the magnitude of the exchange interaction energy, \({\tilde H}_{0}=H_{0}/\gamma |J|\).  We set the scaled exchange interaction energy as \({\tilde J}' \equiv \frac{J}{|J|}=-1\), the scaled cut-off frequency as \({\tilde \omega}_{c}' \equiv \frac{\omega_{c}}{|J|}=0.5\), the scaled coupling strength as \(s=0.02\), and the scaled strength of the dipole interaction as \({\tilde D_{0}}'\equiv \frac{D_{0}}{|J|}=0.1\).  Since the case of  \(\theta _{12}=0\) corresponds to \(H_{0}//c\) and  \(\theta _{12}=\frac{\pi}{2}\) to \(H_{0}\bot c\),  we found that  Fig.\ref{fig:fig11}  shows  the same feature as the resonant shift studied by Nagata and Tazuke.  
 
\section{Born-Markovian Approximation}
\label{sec:c}
Let us show the relation between the formula in this paper and the conventional one in the Born-Markovian approximation.  
First, in the Born approximation, after the transformation of \(s=t-\tau\) on the memory kernel Eq.~(\ref{eqn:a12}), we replace the time evolution of \({\rho_{A}}_{\nu}(t-s)\) as \(e^{i \el_{0} s} {\rho_{A}}_{\nu}(t)\)\cite{hss,Davies}: 
\begin{eqnarray}
\int_{0}^{t} ds \Xi_{2} (s) {\rho_{A}}_{\nu}(t-s) &=& \int_{0}^{t} ds \cp (-i \el_{1}) e^{-i \el_{0} (s)} \cq (-i \el_{1}) e^{i \el_{0} s} {\rho_{A}}_{\nu}(t) \nonumber \\
&=& (-\frac{i}{\hbar})^2 \int_{0}^{t} ds \cp [\ch_{1}, [\ch_{1}(-s),{\rho_{A}}_{\nu}(t)]].
\label{eqn:c1} 
\end{eqnarray} 
Moreover, in the Markovian limit, we assume that the correlation time of the bath is much shorter than that of the relevant system, which means that we make the upper bound of the integral in Eq.~(\ref{eqn:c1}) to be infinity\cite{narrow}.  The inhomogeneous term in Eq.~(\ref{eqn:a7}) can be neglected  in the Markovian limit\cite{Haake}. 

In these approximations, we have the time evolution of \({\rho_{A}}_{\nu}(t)\) in the form
\begin{equation}
\frac{d}{dt} {\rho_{A}}_{\nu}(t)=   - \frac{i}{\hbar} [\cha, {\rho_{A}}_{\nu}(t)] - \int_{0}^{\infty} ds \{ \Phi(s) \hx [\hx(-s), {\rho_{A}}_{\nu}(t)]- \Phi(-s) [\hx(-s), {\rho_{A}}_{\nu}(t)] \hx \},
\label{eqn:c3} 
\end{equation} 
which is written in the Hilbert-Schmidt space as,
\begin{equation}
\frac{d}{dt} \rv(t) = - \frac{i}{\hbar} \chm_{S} \;\rv(t) +
         \chm_{{\rm Markov}} \;\rv(t),
\label{eqn:c4}
\end{equation}
where \(\chm_{{\rm Markov}}\) is given by
\begin{eqnarray}
\chm_{{\rm Markov}} &=&
-\int_{0}^{\infty} d s [ \Phi(s) \{ \hx \hx(-s) \otimes \v1- \hx  \otimes \{\hx(-s)^{\dagger}\}^{*} \}\nonumber \\
&&\hspace{1.5cm} +\Phi(-s)\{ \hx(-s) \otimes \{ \hx^{\dagger}\}^{*} - \v1 \otimes \{\hx^{\dagger} \hx(-s)^{\dagger}\}^{*} \} ]. 
\label{eqn:c5} 
\end{eqnarray}

The complex susceptibility in the Born-Markovian limit is obtained as
\begin{equation}
\chi_{\mu \nu}(\omega)= \frac{i}{\hbar} ({\vec \hb}_{\mu} ,  \;\; \chm_{\chi} \; \rv(0) ).
\label{eqn:c6} 
\end{equation} 
where we denote \(\chm_{\chi}\) as
\begin{equation}
\chm_{\chi}=[i \omega +\frac{i}{\hbar} \chm_{S} -\chm_{{\rm Markov}}]^{-1}.
\label{eqn:c7} 
\end{equation}

For the pure dephasing case in spin-boson model, (\(a=0, c=1\), by setting \(\Lambda=0\) in Eq.~(\ref{eqn:44n})), we obtain 
\begin{eqnarray}
\chm_{{\rm Markov}}=-\frac{{\phi_{4}}_{+}[0,0]}{2}
\left(
\begin{array}{llll}
0&0&0&0\\
0&  1  & 0 & 0\\
0& 0& 1 &  0 \\
 0 & 0&  0 & 0
\end{array}
\right). \nonumber \\
\label{eqn:c8} 
\end{eqnarray} 
The difference of the matrix elements of \(\chm_{{\rm Markov}}\) with those of Eq.~(\ref{eqn:46n}) comes from the  replacement of \({\rho_{A}}_{\nu}(t-s)\) with \(e^{i \el_{0} s} {\rho_{A}}_{\nu}(t)\).

In the Born-Markovian approximation, the transverse susceptibility for pure dephasing case is given by
\begin{equation}
\chi_{+ -}(\omega)= \frac{-2 \tanh{(\beta \hbar \omega_{0})}} { \omega-\omega_{0} - \frac{i}{2} {{\phi_{4}}_{+}[0,0]}}.
\label{eqn:c9} 
\end{equation} 
From the definition of Eqs.~(\ref{eqn:47n}) and (\ref{eqn:49n}), we find that the part of principal value integral cancels in this case.  This means that the frequency shift is not included in the Born-Markovian approximation for the pure dephasing case.

\begin{figure}[h]
\begin{center}
\includegraphics[scale=1.0]{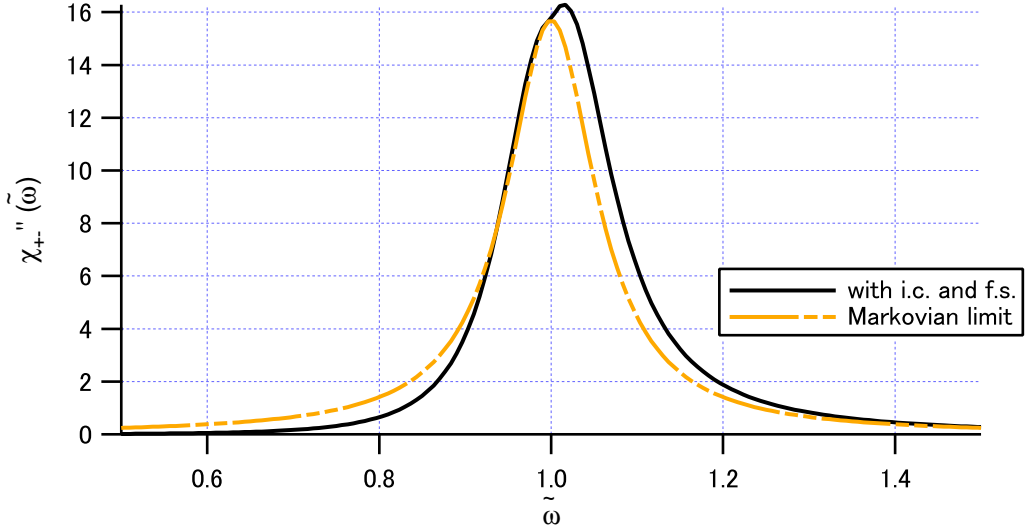}
\end{center}
\caption{(Color Online) Comparison of the transverse susceptibility \(\chi_{+-}''(\omega)\) for \(\Lambda=0\) between the evaluation including the effects of the initial correlation and frequency shift (solid (black) line), and the evaluation in the Markovian limit (double dot-dash (orange) line). The other parameters are the same as in Fig.\ref{fig:fig2}.}
\label{fig:fig11n}
\end{figure}

In Fig.\ref{fig:fig11n}, we compare the transverse susceptibility \(\chi_{+-}''(\omega)\) for \(\Lambda=0\) of the both of the cases Eqs.~(\ref{eqn:49n}) and (\ref{eqn:c9}).  The former includes the effects of the initial correlation and frequency shift (solid (black) line), while the latter is given by the Born-Markovian  approximation(double dot-dash (orange) line).  We find a considerable peak shift which reflects effects of the initial correlation and frequency shift from the Lorentzian line shape in the Born-Markovian approximation.

\begin{thebibliography}{99}
\bibitem{narrow}The procedure of Born-Markovian approximation can also be understood by using the interaction picture and assuming that the time evolution of \({\rho_{A}}_{\nu}(t-s)\) can be approximated with \({\rho_{A}}_{\nu}(t)\) because of the short correlation time of the bath.
\end{thebibliography}
\end{document}